\documentclass[a4paper,12pt]{article}
\usepackage{amssymb}
\usepackage{amsmath}

\newcommand{\be}{\begin{equation}}
\newcommand{\ee}{\end{equation}}
\newcommand{\bea}{\begin{eqnarray}}
\newcommand{\eea}{\end{eqnarray}}
\newcommand{\ba}{\begin{array}}
\newcommand{\ea}{\end{array}}

\makeatletter \@addtoreset{equation}{section} \makeatother

\begin{document}

\begin{titlepage}

    \thispagestyle{empty}
    \begin{flushright}
        \hfill{SU-ITP-10/32}\\
    \end{flushright}

    \vspace{32pt}
    \begin{center}
        { \Huge{\bf Topics in\\Cubic Special Geometry}}

        \vspace{18pt}

        {\large{\bf Stefano Bellucci$^{1}$, Alessio Marrani$^{2}$ and \ Raju Roychowdhury$^{3}$}}

        \vspace{15pt}

        {$1$ \it INFN - Laboratori Nazionali di Frascati, \\
        Via Enrico Fermi 40,00044 Frascati, Italy\\
        \texttt{bellucci@lnf.infn.it}}

        \vspace{15pt}

        {$2$ \it Stanford Institute for Theoretical Physics\\
        Stanford University, Stanford, CA 94305-4060, USA\\
        \texttt{marrani@lnf.infn.it}}

        \vspace{15pt}

        {$3$ \it Dipartimento di Scienze Fisiche, Federico II University,\\
        Complesso Universitario di Monte S. Angelo,\\
        Via Cinthia, Ed. 6, I-80126 Napoli, Italy\\
        \texttt{raju@na.infn.it}}

\end{center}

\vspace{25pt}

\begin{abstract}
We reconsider the sub-leading quantum perturbative corrections to $\mathcal{N%
}=2$ cubic special K\"{a}hler geometries. Imposing the invariance
under axion-shifts, all such corrections (but the imaginary constant
one) can be introduced or removed through suitable, lower
unitriangular symplectic transformations, dubbed Peccei-Quinn (PQ)
transformations.

Since PQ transformations do not belong to the $d=4$ $U$-duality group $G_{4}$%
, in symmetric cases they generally have a non-trivial action on the
unique
quartic invariant polynomial $\mathcal{I}_{4}$ of the charge representation $%
\mathbf{R}$ of $G_{4}$. This leads to interesting phenomena in
relation to theory of extremal black hole attractors; namely, the
possibility to make transitions between different charge orbits of
$\mathbf{R}$, with corresponding change of the supersymmetry
properties of the supported attractor solutions. Furthermore, a
suitable action of PQ transformations can also set $\mathcal{I}_{4}$
to zero, or vice versa it can generate a non-vanishing
$\mathcal{I}_{4}$: this corresponds to transitions between ``large''
and ``small'' charge orbits, which we classify in some detail within
the ``special coordinates'' symplectic frame.

Finally, after a brief account of the action of PQ transformations
on the recently established correspondence between Cayley's
hyperdeterminant and
elliptic curves, we derive an equivalent, alternative expression of $%
\mathcal{I}_{4}$, with relevant application to black hole entropy.

\end{abstract}

\end{titlepage}
\tableofcontents
\section{Introduction}

Special K\"{a}hler geometry (SK) characterizes the scalar manifolds of
Abelian vector multiplets in $\mathcal{N}=2$ supergravity theory in $d=4$
space-time dimensions (see \textit{e.g.} \cite{CDF-rev,N=2-big,Craps,Freed},
and Refs. therein). Along the years, it has played a key role in various
important developments in black hole (BH) physics.

Among these, the \textit{Attractor Mechanism} \cite{AM-Refs} shed light on
the dynamics of scalar fields coupled to BPS (Bogomol'ny-Prasad-Sommerfeld)
and non-BPS extremal BHs. Through the introduction of an effective BH
potential $V_{BH}$ \cite{FGK}, this mechanism describes the stabilization of
the scalar fields in terms of the BH conserved charges in the near-horizon
limit of the extremal BH background (see \textit{e.g.} \cite
{ADFT-rev,Sen-EF,Sen-rev,Kallosh-rev,FHM-rev,BFGM2}, also for reviews and
lists of Refs.).

Within theories with $\mathcal{N}=2$ local supersymmetry emerging from
Calabi-Yau compactifications of superstrings or $M$-theory, the Attractor
Mechanism has played a key role in the study of connections with topological
string partition functions \cite{OSV} and relations with microstates
counting (see for instance \cite{Sen-rev}), and also in the investigation of
dynamical phenomena, such as wall crossing and split attractor flow (see
\textit{e.g.} \cite{MS-Refs}, and Refs. therein).

In some seminal papers dating back to mid 90's \cite{AM-Refs}, the Attractor
Mechanism was discovered by Ferrara, Kallosh and Strominger in $\mathcal{N}%
=2 $, $d=4$ \textit{ungauged} supergravity coupled to $n_{V}$ vector
multiplets. This theory proved to be an especially relevant and rich
framework for the study of the attractor dynamics of scalar flows coupled to
extremal BHs.

An important arena in which many advances have been made along the years is
provided by a particular yet broad class of SK geometries, namely the ones
determined by an holomorphic prepotential function $F$ which is purely cubic
in the complex scalar fields themselves:
\begin{equation}
\mathcal{F}_{d}\equiv \frac{1}{3!}d_{ijk}z^{i}z^{j}z^{k}.  \label{d-SKG-pre}
\end{equation}
$\mathcal{F}_{d}$ defines the so-called $d$-SK geometries \cite{dWVP-2,dWVVP}%
. These geometries naturally arise as the large volume limit of $CY_{3}$
compactifications of Type II(A) superstring theories, in which $d_{ijk}$ is
given by the triple intersection numbers of the $CY_{3}$ internal manifold
itself (see Sec. \ref{Stringy} for further details, and list of Refs.).

Moreover, up to the so-called \textit{minimal coupling} sequence (with
quadratic prepotential) \cite{Luciani}, all non-compact \textit{symmetric}
coset SK spaces $\frac{G_{4}}{H_{4}}$ are actually $d$-spaces, defined by a
prepotential of the form \cite{dWVVP}; $G_{4}$ is the $d=4$ $U$-duality group%
\footnote{%
Here $U$-duality is referred to as the ``continuous'' limit (valid for large
values of the charges) of the non-perturbative string theory symmetries
introduced by Hull and Townsend in \cite{HT}.}, and $H_{4}$ is its maximal
compact subgroup (with symmetric embedding). In symmetric SK geometries the
Attractor Mechanism enjoys a noteworthy geometrical interpretation, related
to the fascinating interplay among orbits of the charge irrepr. $\mathbf{R}$
of $G_{4}$ \cite{FG1,BFGM1}, the solution of the Attractor Eqs. \cite{BFGM1}
and the related \textit{``moduli spaces''} \cite{Ferrara-Marrani-2}. Through
the Bekenstein-Hawking entropy ($S$) -area ($A$) formula \cite{BH1}
\begin{equation}
\frac{S}{\pi }=\frac{A}{4}=\sqrt{\left| \mathcal{I}_{4}\left( \mathcal{Q}%
\right) \right| },  \label{Bek}
\end{equation}
the entropy of the BH is given in terms of the unique invariant polynomial $%
\mathcal{I}_{4}$ of the charge irrepr. $\mathbf{R}$ of $G_{4}$, which is
quartic in charges $\mathcal{Q}$. It is also worth recalling that also the
recently introduced first order approach to non-BPS scalar flows \cite{CD}
has been completely solved in terms of geometrical quantities ($U$-duality
invariants) in \cite{FO-Refs}.

It is therefore natural to ask what is the role and the effect of
sub-leading corrections to the $\mathcal{N}=2$ purely cubic prepotential (%
\ref{d-SKG-pre}). As it is well known (see the recent discussion in \cite
{Raju-1}, and Refs. therein), such corrections are of both quantum
perturbative and non-perturbative nature, and not all of them are consistent
with the Peccei-Quinn axion-shift symmetry \cite{Peccei-Quinn}, nor all of
them actually affect the SK geometry of the scalar manifold itself (see
\textit{e.g.} \cite{CFG}).

In this paper, extending on some previous results in \cite
{dWVP-2,N=2-Quantum,Shmakova}, we further develop the study of those
sub-leading corrections to $d$-SK geometries (\ref{d-SKG-pre}) which are
consistent with the axion-shift symmetry and which do not affect the
geometry of the vector multiplets' scalar fields\footnote{%
For a recent discussion of the unique (constant imaginary) term which is
consistent with axion-shift and affects the geometry, see \textit{e.g.} \cite
{Raju-1}.}.

It is known \cite{dWVP-2,N=2-Quantum} that these sub-leading corrections can
be included in (or removed from) the $\mathcal{N}=2$ symplectic sections by
acting with suitable symplectic transformations, and this provides an
effective shortcut to the process of solving the Attractor Eqs. (\textit{%
alias} criticality conditions for $V_{BH}$) in the so corrected $d$-SK
geometries. As we will find in the present investigation, such symplectic
transformations have a group structure (we dub them Peccei-Quinn (PQ)
symplectic transformations), but they do not belong to the suitable
symplectic representation of $G_{4}$ itself.

At least for symmetric $d$-SK geometries, this leads to interesting
consequences in the theory of charge orbits and \textit{``moduli spaces''}
of extremal BH attractor solutions. Indeed, the PQ transformations do not
affect the geometry of the scalar manifold, neither the statification of the
charge irrepr. space $\mathbf{R}$ into disjoint orbits, nor the structure of
the corresponding \textit{``moduli spaces''} of attractors\footnote{%
In this respect, the general analysis and findings of the present paper
explains the result obtained in Sec. 3 and App. A of \cite{DT-1}, also
providing a way to generalise them to generic BH charge configuration, and
to a generic model with $n_{V}$ vector multiplets.
\par
Moreover, through the action of PQ symplectic group, also the results
concerning non-perturbative instantonic corrections to the prepotential,
obtained in Sec. 4 and App. B of \cite{DT-1}, can be generalised to include
the sub-leading quantum perturbative corrections under consideration. See
treatment below for further comments.}, but they can change the value and
the sign of $\mathcal{I}_{4}$, thus possibly switching from one charge
orbits to another.

For instance, an extremal ``small'' BH configuration (with vanishing entropy
according to formula (\ref{Bek})) within the $d$-SK geometry (\ref{d-SKG-pre}%
) can acquire, by introducing the quantum perturbative correction under
consideration, a non-vanishing area of the event horizon, and thus a
``large'' nature (namely, a non-vanishing $\mathcal{I}_{4}$, and thus
entropy, according to (\ref{Bek})). The opposite phenomenon can occur too,\
namely that ``large'' extremal BH configuration can become ``small'' for
particular choices of the supporting charge vectors.

Another possible phenomenon is that the supersymmetry preserving features of
the attractor configurations of $d$-SK geometry (\ref{d-SKG-pre}) can change
in presence of those sub-leading corrections accounted for by PQ
transformations. This is somewhat analogous to some phenomena observed in
presence of the ``$+i\lambda $'' correction in the prepotential in \cite
{BFMS}.

By exploiting the PQ symplectic transformation, we will also study how the
effective BH potential $V_{BH}$ gets modified in presence of the
aforementioned corrections, and what is the fate of those charge
configurations which support axion-free attractor solutions within the
theory determined by (\ref{d-SKG-pre}). In general, the solutions of
Attractor Eqs. for the corrected d-SK geometries can be obtained by
considering the solutions in the purely cubic theory \cite{Shmakova,TT1},
and by transforming the charges in such formul\ae\ with a suitable PQ
transformation.

We will also briefly comment on the action of the PQ group on the roots of
certain cubic elliptic curves, which have been recently connected \cite{G-1}
to the Cayley's hyperdeterminant \cite{Cayley}, namely to the (opposite of) $%
\mathcal{I}_{4}$ for the noteworthy \textit{triality-symmetric} so-called $%
stu$ supergravity model \cite{stu}. This might lead to an interpretation of
the PQ transformation within the intriguing ``BH/qubit correspondence'' \cite
{QIT}.

Finally, we derive an alternative expression of $\mathcal{I}_{4}$ for
symmetric $d$-SK geometries, and more in general for symmetric cubic
geometries (such as the ones of some $\mathcal{N}>2$-extended, $d=4$
supergravities). This result allows for a consistent treatment of some
expressions of the BH entropy available in the literature (see \textit{e.g.}
\cite{TT1}). Furthermore, its further generalisation to the case of
non-symmetric geometries (in which $\mathcal{I}_{4}$ is not generally
related to the BH entropy) explicitly shows the contribution of the
so-called $E$-tensor \cite{dWVVP} introducing an explicit dependence on
(some of the) scalar degrees of freedom.\bigskip\

The plan of the paper is as follows.

In Sec. \ref{GT} we analyse the PQ symplectic transformations within $%
\mathcal{N}=2$, $d=4$ SK geometry. More specifically, in Sec. \ref
{Axion-Shift} we recall the general structure of sub-leading terms in cubic
prepotential, and their consistency with axion-shift symmetry. The PQ
symplectic group is introduced in Sec. \ref{PQ-Group}, and its relation to
the $U$-duality group clarified in Sec. \ref{Rel-U-Duality}. Moreover, Sec.
\ref{Stringy} considers some aspects of stringy origin and topological
interpretation of some generators of the PQ group.

Then, Sec. \ref{Applications} applies this general formalism to relevant
issues within the theory of extremal black hole attractors. Secs. \ref
{Transformation-I4} and \ref{Analysis-Configs} and is devoted to the study
and classification (within symmetric cubic geometries) of the PQ group on
the unique invariant polynomial $\mathcal{I}_{4}$ of the charge
representation $\mathbf{R}$ of the $U$-duality group. At the end of Sec. \ref
{Analysis-Configs}, we briefly comment on the relevance of the PQ group for
the attractor values of the scalars, \textit{i.e.} for the non-degenerate
critical points of the effective BH potential $V_{BH}$. The transformation
properties of the latter are studied in Sec. \ref{VBH-Transf}, with an
analysis of the possible axion-free supporting charge configurations.

Sec. \ref{PQ-Ell-Cayley} briefly analyses the \textit{``PQ-deformation''} of
the recently established intriguing relation between Cayley's
hyperdeterminant and elliptic curves.

Finally, in Sec. \ref{Alternative-I4} an equivalent, alternative expression
for $\mathcal{I}_{4}$ is derived, by exploiting the identities
characterising symmetric cubic special geometries, with relevant
consequences on the matching of known expressions of the black hole entropy.
In particular, the new expression $\mathcal{I}_{4}$ allows one to relate its
the scalar-dependence in non-symmetric geometries directly to the so-called $%
E$-tensor.

\section{\label{Peccei-Quinn}Peccei-Quinn Symplectic Transformations}

\subsection{\label{GT}General Theory}

Let us consider $\mathcal{N}=2$, $d=4$ \textit{ungauged} Maxwell-Einstein
supergravity, whose vector multiplets' scalar manifold is endowed with
special K\"{a}hler (SK) geometry, based on an holomorphic prepotential
function $F$, homogeneous of degree $2$ in the contravariant symplectic
sections $X^{\Lambda }$ (the reader is addressed \textit{e.g.} to \cite
{CDF-rev,N=2-big,Craps,Freed} for a thorough introduction and list of Refs.).

\subsubsection{\label{Axion-Shift}Cubic Special Geometries and Axion-Shifts}

We start and define the most general form of \textit{cubic} prepotential as
follows\footnote{%
Greek capital and Latin lowercase indices respectively run $0,1,...,n_{V}$
and $1,...,n_{V}$ throughout. The naught index pertains to the graviphoton,
while $n_{V}$ denotes the number of Abelian vector multiplets coupled to the
supergravity one. Therefore, we work within the so-called symplectic basis
of \textit{special coordinates} (see e.g. \cite{dWVVP,N=2-big} and Refs.
therein), which is manifestly covariant with respect to the $d=5$ $U$%
-duality group $G_{5}$.} ($d_{\Lambda \Sigma \Xi }=d_{\left( \Lambda \Sigma
\Xi \right) }\in \mathbb{C}$):
\begin{eqnarray}
F &\equiv &\frac{1}{3!}d_{\Lambda \Sigma \Xi }\frac{X^{\Lambda }X^{\Sigma
}X^{\Xi }}{X^{0}}=  \notag \\
&=&\frac{1}{3!}\left( \text{Re}d_{ijk}+i\text{Im}d_{ijk}\right) \frac{%
X^{i}X^{j}X^{k}}{X^{0}}+\frac{1}{2}\left( \text{Re}d_{0ij}+i\text{Im}%
d_{0ij}\right) X^{i}X^{j}+  \notag \\
&&+\frac{1}{2}\left( \text{Re}d_{00i}+i\text{Im}d_{00i}\right) X^{i}X^{0}+%
\frac{1}{3!}\left( \text{Re}d_{000}+i\text{Im}d_{000}\right) \left(
X^{0}\right) ^{2}.  \label{F-cubic}
\end{eqnarray}
By denoting the real and imaginary part of $X^{i}$ respectively as $%
X^{i}\equiv R^{i}+iI^{i}$, the corresponding K\"{a}hler potential reads%
\footnote{%
For simplicity's sake, in Eqs. (\ref{K}), (\ref{axion-shift}) and (\ref{K-2}%
) we give the result for $X^{0}\equiv 1$, which does not imply any loss of
generality for our purposes.}
\begin{eqnarray}
\mathcal{K} &\equiv &-\ln \left[ i\left( X^{\Lambda }\overline{F}_{\Lambda }-%
\overline{X}^{\Lambda }F_{\Lambda }\right) \right]  \notag \\
&=&-\frac{4}{3}i\text{Re}d_{ijk}I^{i}I^{j}I^{k}-\frac{2}{3}i\text{Im}%
d_{ijk}R^{i}R^{j}R^{k}-2i\text{Im}d_{ijk}R^{i}I^{j}I^{k}  \notag \\
&&-2i\text{Im}d_{0ij}R^{i}R^{j}-2i\text{Im}d_{0ij}I^{i}I^{j}-2i\text{Im}%
d_{00i}R^{i}-\frac{2}{3}i\text{Im}d_{000}.  \label{K}
\end{eqnarray}
Thus, the invariance of $\mathcal{K}$ under Peccei-Quinn (PQ) perturbative
(continuous) axion-shift symmetry \cite{Peccei-Quinn}
\begin{equation}
R^{i}\rightarrow R^{i}+\alpha ^{i},~\alpha ^{i}\in \mathbb{R}
\label{axion-shift}
\end{equation}
yields
\begin{equation}
\text{Im}d_{ijk}=\text{Im}d_{0ij}=\text{Im}d_{00i}=0.
\end{equation}
The resulting axion-shift-invariant expression of $\mathcal{K}$ then simply
reads
\begin{equation}
\mathcal{K}=-\frac{4}{3}i\text{Re}d_{ijk}I^{i}I^{j}I^{k}-\frac{2}{3}i\text{Im%
}d_{000},  \label{K-2}
\end{equation}
and the prepotential $F$ given by (\ref{F-cubic}) can accordingly be split
as
\begin{equation}
F=\mathbf{F}+\frak{F},  \label{F-split}
\end{equation}
where
\begin{equation}
\mathbf{F}\equiv \frac{1}{3!}\text{Re}d_{ijk}\frac{X^{i}X^{j}X^{k}}{X^{0}}+%
\frac{i}{3!}\text{Im}d_{000}\left( X^{0}\right) ^{2}  \label{F-bold}
\end{equation}
is the part contributing to $\mathcal{K}$ given by (\ref{K-2}) and thus to
the SK geometry, and
\begin{equation}
\frak{F}\equiv \frac{1}{2}\text{Re}d_{0ij}X^{i}X^{j}+\frac{1}{2}\text{Re}%
d_{00i}X^{i}X^{0}+\frac{1}{3!}\text{Re}d_{000}\left( X^{0}\right) ^{2}
\label{F-sympl}
\end{equation}
is a \textit{quadratic} form in $X^{\Lambda }$, which does not contribute to
$\mathcal{K}$. Thus, $\mathbf{F}$ given by (\ref{F-bold}) is the most
general cubic prepotential which is consistent with the PQ axion-shift (\ref
{axion-shift}) and which affects the geometry of the scalar manifold itself
\cite{CFG}. Some issues within the SK geometry based on $\mathbf{F}$ have
been recently investigated in \cite{Raju-1} (see also \cite{N=2-Quantum}).

On the other hand, Re$d_{ijk}$ is usually denoted simply by the real symbol $%
d_{ijk}$, and the holomorphic function
\begin{equation}
\mathbf{F}_{d}\equiv \frac{1}{3!}d_{ijk}\frac{X^{i}X^{j}X^{k}}{X^{0}}
\label{d-SKG}
\end{equation}
is the prepotential of the so-called $d$-SK geometries\footnote{%
Regardless of the explicit form of $d_{ijk}$, the corresponding special
K\"{a}hler manifold has always \textit{at least }$n_{V}+1$ global
isometries, namely an overall scaling and PQ axion-shifts (see Eq. (\ref
{axion-shift})), forming the group $SO\left( 1,1\right) \times _{s}\mathbb{R}%
^{n_{V}}$, which can be considered the ``minimal $G_{4}$'' of $d$-SK
geometries. Its relation to $d=5$ uplift and further details can be found
\textit{e.g.} in \cite{ADFT-flat-1} (see also Refs. therein).} \cite
{dWVP-1,dWVVP}. This will be the most general framework we will be
considering in the applications of Sec. \ref{Applications}.

For later convenience, let us compute the derivatives of $\frak{F}$ with
respect to the sections\footnote{%
As shown in \cite{Strominger-SKG}, the symplectic connection of SK geometry
is \textit{flat}.} $X^{\Lambda }$:
\begin{equation}
\frak{F}_{\Lambda }\equiv D_{\Lambda }\frak{F}=\frac{\partial \frak{F}}{%
\partial X^{\Lambda }}=\left\{
\begin{array}{l}
\frak{F}_{0}=\frac{1}{2}\text{Re}d_{00i}X^{i}+\frac{1}{3}\text{Re}%
d_{000}X^{0}; \\
\\
\frak{F}_{i}=\text{Re}d_{0ij}X^{j}+\frac{1}{2}\text{Re}d_{00i}X^{0}.
\end{array}
\right.  \label{FfrakL}
\end{equation}
\medskip\

It has been known (see \textit{e.g.} \cite{dWVP-2,N=2-Quantum,Shmakova})
that $\frak{F}$ can be introduced (or removed) in any $\mathcal{N}=2$
prepotential by performing suitable symplectic transformations. More
specifically, through the action of particular symplectic transformations
one can introduce the effect of the sub-leading quantum perturbative terms (%
\ref{F-sympl}) into the explicit expression of horizon values of attractors
and into the corresponding value of BH entropy \cite{N=2-Quantum,Shmakova}.

A major part of the present investigation is devoted to a thorough analysis
of this issue in full generality. In particular, we will focus on the effect
of $\frak{F}$ on the BH entropy in the general framework of $d$-SKG, with
leading cubic prepotential given by (\ref{d-SKG}). This will naturally lead
to the study of the effect of the so-called \textit{Peccei-Quinn
transformations}, namely particular symplectic transformations deeply
related to to $\frak{F}$, on the duality invariants and supersymmetry
properties of extremal BH attractor solutions.

The results recently obtained in Sec. 3 of \cite{DT-1} provide an explicit
example (with $n_{V}=2$ and for a particular charge configuration) of some
aspects of the general treatment given here. Indeed, the prepotential given
by Eq. (3.7) of \cite{DT-1} is nothing but a particular case\footnote{%
In this respect (and referring to the equation numbering of \cite{DT-1}), it
is worth noting that the second of Eqs. (3.8) can be directly obtained from
the general expression (2.9) for $d$-SK geometry, because the sub-leading
quantum perturbative terms appearing in Eq. (3.7) do \textit{not} affect the
K\"{a}hler potential and thus the metric.} of the general structure (\ref
{F-split})-(\ref{F-sympl}).

\subsubsection{\label{PQ-Group}The Peccei-Quinn Symplectic Group}

Given an element\footnote{%
In all the following treatment, we work in the \textit{(semi)classical}
limit of large (continuous) charges, thus the field of definition of
considered linear and symplectic groups is $\mathbb{R}$, and not $\mathbb{Z}$%
, as instead it would pertain to the \textit{quantum} level.}
\begin{equation}
\mathcal{S}\equiv \left(
\begin{array}{cc}
\mathcal{U} & \mathcal{Z} \\
\mathcal{W} & \mathcal{V}
\end{array}
\right) \in GL\left( 2n_{V}+2,\mathbb{R}\right) ,  \label{S-call}
\end{equation}
it belongs to the symplectic group $Sp\left( 2n_{V}+2,\mathbb{R}\right)
\subsetneq GL\left( 2n_{V}+2,\mathbb{R}\right) $ \textit{iff}
\begin{equation}
\mathcal{S}^{T}\Omega \mathcal{S}=\Omega \Leftrightarrow \mathcal{S}%
^{-1}=\Omega ^{-1}\mathcal{S}^{T}\Omega =-\Omega \mathcal{S}^{T}\Omega ,
\label{finite-sympl-cond}
\end{equation}
where $\Omega $ is the $\left( 2n_{V}+2\right) \times \left( 2n_{V}+2\right)
$ symplectic metric (the subscripts denote the dimensions of the square
block components):
\begin{equation}
\Omega \equiv \left(
\begin{array}{cc}
0_{n_{V}+1} & \mathbb{I}_{n_{V}+1} \\
-\mathbb{I}_{n_{V}+1} & 0_{n_{V}+1}
\end{array}
\right) .
\end{equation}
The finite condition of symplecticity (\ref{finite-sympl-cond}) translates
on the square block components of $\mathcal{S}$ as follows:
\begin{eqnarray}
\mathcal{U}^{T}\mathcal{V}-\mathcal{W}^{T}\mathcal{Z} &=&\mathbb{I}%
_{n_{V}+1};  \label{sympl-cond-1} \\
\mathcal{U}^{T}\mathcal{W}-\mathcal{W}^{T}\mathcal{U} &=&\mathcal{Z}^{T}%
\mathcal{V}-\mathcal{V}^{T}\mathcal{Z}=0_{n_{V}+1}.  \label{sympl-cond-2}
\end{eqnarray}

In general, the $U$-duality group $G_{4}$ of $\mathcal{N}=2$, $d=4$
supergravity is embedded into $Sp\left( 2n_{V}+2,\mathbb{R}\right) $ through
its relevant (namely, smallest symplectic) (ir)repr. $\mathbf{R}$ (see
\textit{e.g.} \cite{N=2-big} and Refs. therein):
\begin{equation}
G_{4}\overset{\mathbf{R}}{\subsetneq }Sp\left( 2n_{V}+2,\mathbb{R}\right) .
\label{G-sympl-embed}
\end{equation}
The vector of the fluxes of the two-form field strengths of the Abelian
vector fields and of their duals
\begin{equation}
\mathcal{Q}\equiv \left( p^{\Lambda },q_{\Lambda }\right) ^{T}=\left(
p^{0},p^{i},q_{0},q_{i}\right) ^{T},
\end{equation}
as well as the vector of the holomorphic sections
\begin{equation}
\mathbf{V}\equiv \left( X^{\Lambda },F_{\Lambda }\right) ^{T}=\left(
X^{0},X^{i},F_{0},F_{i}\right) ^{T},  \label{V-bold}
\end{equation}
sit in $\mathbf{R}$, and thus they are $Sp\left( 2n_{V}+2,\mathbb{R}\right) $%
-covariant, transforming under $\mathcal{S}$ as follows:
\begin{eqnarray}
\mathcal{Q}^{\prime } &=&\mathcal{SQ}=\left(
\begin{array}{c}
\mathcal{U}_{~\Sigma }^{\Lambda }p^{\Sigma }+\mathcal{Z}^{\Lambda \Sigma
}q_{\Sigma } \\
\mathcal{W}_{\Lambda \Sigma }p^{\Sigma }+\mathcal{V}_{\Lambda }^{~\Sigma
}q_{\Sigma }
\end{array}
\right) ;  \label{charge-sympl} \\
&&  \notag \\
\mathbf{V}^{\prime } &=&\mathcal{S}\mathbf{V}=\left(
\begin{array}{c}
\mathcal{U}_{~\Sigma }^{\Lambda }X^{\Sigma }+\mathcal{Z}^{\Lambda \Sigma
}F_{\Sigma } \\
\mathcal{W}_{\Lambda \Sigma }X^{\Sigma }+\mathcal{V}_{\Lambda }^{~\Sigma
}F_{\Sigma }
\end{array}
\right) .  \label{section-sympl}
\end{eqnarray}
\medskip

Now, by recalling (\ref{FfrakL}), it is immediate to realize that $\frak{F}%
_{\Lambda }$ can be generated or removed by performing a suitable symplectic
finite transformation on $\mathbf{V}$. Indeed, the identification
\begin{equation}
\frak{F}_{\Lambda }\equiv F_{\Lambda }^{\prime }-\mathcal{V}_{\Lambda
}^{~\Sigma }F_{\Sigma }=\mathcal{W}_{\Lambda \Sigma }X^{\Sigma }=\mathcal{W}%
_{\Lambda 0}X^{0}+\mathcal{W}_{\Lambda i}X^{i}  \label{def1}
\end{equation}
defines, through Eq. (\ref{section-sympl}), the components of the $\left(
n_{V}+1\right) \times \left( n_{V}+1\right) $ sub-matrix $\mathcal{W}%
_{\Lambda \Sigma }$:
\begin{equation}
\mathcal{W}_{\Lambda \Sigma }=\left(
\begin{array}{ccc}
\mathcal{W}_{00} &  & \mathcal{W}_{0j} \\
&  &  \\
\mathcal{W}_{i0} &  & \mathcal{W}_{ij}
\end{array}
\right) \equiv \frac{1}{3!}\left(
\begin{array}{ccc}
2\text{Re}d_{000} &  & 3\text{Re}d_{00j} \\
&  &  \\
3\text{Re}d_{00i} &  & 6\text{Re}d_{0ij}
\end{array}
\right) \equiv \left(
\begin{array}{ccc}
\varrho &  & \mathbf{c}_{j} \\
&  &  \\
\mathbf{c}_{i} &  & \Theta _{ij}
\end{array}
\right) =\mathcal{W}_{\left( \Lambda \Sigma \right) },  \label{def2}
\end{equation}
which inherits the symmetry properties from the relevant components of the $%
d_{\Lambda \Sigma \Xi }$ tensor. Note that we re-named the quantities for
simplicity's sake ($\Theta _{ij}=\Theta _{\left( ij\right) }$).

Thus, we are going to deal with particular symplectic transformations
defined as follows:

\begin{enumerate}
\item  In order to keep the contravariant symplectic sections $X^{\Lambda }$
(and thus the coordinates of the scalar manifold) \textit{invariant} under
the considered transformations, (\ref{section-sympl}) imposes
\begin{equation}
\mathcal{Z}^{\Lambda \Sigma }\equiv 0,~\mathcal{U}_{~\Sigma }^{\Lambda
}\equiv \delta _{\Sigma }^{\Lambda }.  \label{def4}
\end{equation}

\item  In order to generate or remove $\frak{F}_{\Lambda }$, as stated above
one must define $\mathcal{W}_{\Lambda \Sigma }$ as in Eq. (\ref{def2}), and
furthermore Eq. (\ref{section-sympl}) yields
\begin{equation}
\mathcal{V}_{\Lambda }^{~\Sigma }\equiv \delta _{\Sigma }^{\Lambda }.
\label{def3}
\end{equation}
\end{enumerate}

The $\left( n_{V}+1\right) \times \left( n_{V}+1\right) $ matrices $\mathcal{%
U}$, $\mathcal{Z}$, $\mathcal{V}$ and $\mathcal{W}$ defined by Eqs. (\ref
{def2}), (\ref{def3}) and (\ref{def4}) do satisfy the finite symplecticity
condition (\ref{finite-sympl-cond}), and we denote the corresponding
symplectic matrix as
\begin{equation}
\mathcal{O}\equiv \left(
\begin{array}{ccc}
\mathbb{I}_{n_{V}+1} &  & 0_{n_{V}+1} \\
&  &  \\
\mathcal{W} &  & \mathbb{I}_{n_{V}+1}
\end{array}
\right) .  \label{O}
\end{equation}

It is easy to realize that $\mathcal{O}$ given by (\ref{O}) belongs to the $%
\frac{\left( n_{V}+1\right) \left( n_{V}+2\right) }{2}$-dimensional Abelian
group
\begin{equation}
\mathcal{PQ}\left( 2n_{V}+2,\mathbb{R}\right) \equiv Sp\left( 2n_{V}+2,%
\mathbb{R}\right) \cap LUT\left( 2n_{V}+2,\mathbb{R}\right) ,  \label{rel-1}
\end{equation}
which we will henceforth refer to as the \textit{Peccei Quinn symplectic
group}. In (\ref{rel-1})\linebreak\ $LUT\left( 2n_{V}+2,\mathbb{R}\right) $
is the $\left( n_{V}+1\right) ^{2}$-dimensional Abelian group of lower
unitriangular\linebreak\ $2\left( n_{V}+1\right) \times 2\left(
n_{V}+1\right) $ real matrices, which are unipotent (see \textit{e.g.} \cite
{Unip-1}). Correspondingly, the Peccei-Quinn (PQ) symplectic Lie algebra $%
\frak{pq}\left( 2n_{V}+2,\mathbb{R}\right) $ is given by
\begin{equation}
\frak{pq}\left( 2n_{V}+2,\mathbb{R}\right) \equiv \frak{sp}\left( 2n_{V}+2,%
\mathbb{R}\right) \cap \frak{lut}\left( 2n_{V}+2,\mathbb{R}\right) ,
\label{rel-2}
\end{equation}
namely by the strictly lower triangular $2\left( n_{V}+1\right) \times
2\left( n_{V}+1\right) $ real matrices (which are nilpotent) with \textit{%
symmetric} lower $\left( n_{V}+1\right) \times \left( n_{V}+1\right) $
block.\medskip

Matrices with structure as $\mathcal{O}$ given by (\ref{O}), and thus
belonging to the group\linebreak\ $\mathcal{PQ}\left( 2n_{V}+2,\mathbb{R}%
\right) $ defined above, appear also in other contexts. For instance, they
are a particular case (with $A=\mathbb{I}_{n_{V}+1}$) of the quantum
perturbative duality transformations in supersymmetric Yang-Mills theories
coupled to supergravity (see \textit{e.g.} \cite{CDFVP-1}, and Eq. (4.1)
therein). In particular, Eq. (\ref{O}) defines the structure of quantum
perturbative monodromy matrices in heterotic string compactifications with
classical $U$-duality $SL\left( 2,\mathbb{R}\right) \times SO\left(
2,n_{V}+2\right) $ (see \textit{e.g.} (5.4) of \cite{CDFVP-1}).\medskip

Let us give here some other explicit results, useful in the subsequent
treatment.

Eqs. (\ref{def1}, (\ref{def2}) and (\ref{def3})\ imply
\begin{equation}
\frak{F}_{\Lambda }\equiv F_{\Lambda }^{\prime }-F_{\Lambda }.
\end{equation}
Thus, within the framework under consideration, it follows that
\begin{eqnarray}
F_{\Lambda } &\equiv &D_{\Lambda }\mathbf{F}=\frac{\partial \mathbf{F}}{%
\partial X^{\Lambda }}=\left\{
\begin{array}{l}
F_{0}=-\frac{1}{3!}\text{Re}d_{ijk}\frac{X^{i}X^{j}X^{k}}{\left(
X^{0}\right) ^{2}}+\frac{i}{3}\text{Im}d_{000}X^{0}; \\
\\
F_{i}=\frac{1}{2}\text{Re}d_{ijk}\frac{X^{j}X^{k}}{X^{0}};
\end{array}
\right. \\
F_{\Lambda }^{\prime } &\equiv &D_{\Lambda }\frak{F}+D_{\Lambda }\mathbf{F}%
=D_{\Lambda }F=\frac{\partial F}{\partial X^{\Lambda }},
\end{eqnarray}
where Eqs. (\ref{F-split}) and (\ref{F-bold}) were used.

Moreover, by using (\ref{finite-sympl-cond}), the inverse of matrix $%
\mathcal{O}$ can be easily computed to be simply
\begin{equation}
\mathcal{O}^{-1}\equiv \left(
\begin{array}{ccc}
\mathbb{I}_{n_{V}+1} &  & 0_{n_{V}+1} \\
&  &  \\
-\mathcal{W} &  & \mathbb{I}_{n_{V}+1}
\end{array}
\right) .  \label{O-inv}
\end{equation}
Thus, by recalling Eqs. (\ref{charge-sympl}), (\ref{section-sympl}), and the
expressions (\ref{O}) and (\ref{O-inv}) along with Eq. (\ref{def2}), one can
write down the finite transformations of $\mathcal{Q}$ and $\mathbf{V}$
under the action of a generic element of $\mathcal{PQ}\left( 2n_{V}+2,%
\mathbb{R}\right) $ (the unwritten matrix components vanish throughout):
\begin{eqnarray}
\mathcal{Q}^{\prime } &=&\mathcal{OQ}=\left(
\begin{array}{l}
p^{0} \\
p^{i} \\
q_{0}+\varrho p^{0}+\mathbf{c}_{j}p^{j} \\
q_{i}+\mathbf{c}_{i}p^{0}+\Theta _{ij}p^{j}
\end{array}
\right) \Leftrightarrow \mathcal{Q}=\mathcal{O}^{-1}\mathcal{Q}^{\prime
}=\left(
\begin{array}{l}
p^{\prime 0} \\
p^{\prime i} \\
q_{0}^{\prime }-\varrho p^{\prime 0}-\mathbf{c}_{j}p^{\prime j} \\
q_{i}^{\prime }-\mathbf{c}_{i}p^{\prime 0}-\Theta _{ij}p^{\prime j}
\end{array}
\right) ;  \label{PQ-charge} \\
&&  \notag \\
\mathbf{V}^{\prime } &=&\mathcal{O}\mathbf{V}=\left(
\begin{array}{l}
X^{0} \\
X^{i} \\
F_{0}+\varrho X^{0}+\mathbf{c}_{j}X^{j} \\
F_{i}+\mathbf{c}_{i}X^{0}+\Theta _{ij}X^{j}
\end{array}
\right) \Leftrightarrow \mathbf{V}=\mathcal{O}^{-1}\mathbf{V}^{\prime
}=\left(
\begin{array}{l}
X^{\prime 0} \\
X^{\prime i} \\
F_{0}^{\prime }-\varrho X^{\prime 0}-\mathbf{c}_{j}X^{\prime j} \\
F_{i}^{\prime }-\mathbf{c}_{i}X^{\prime 0}-\Theta _{ij}X^{\prime j}
\end{array}
\right) .  \notag \\
&&  \label{PQ-section}
\end{eqnarray}

\subsubsection{\label{Rel-U-Duality}Relation with $U$-Duality Transformations%
}

In order to highlight some important features of the Peccei-Quinn
transformations defined above, it is here convenient to briefly recall the
properties of $\mathbf{V}$ and related quantities under the action of $%
Sp\left( 2n_{V}+2,\mathbb{R}\right) $ (see \textit{e.g.} \cite
{Fré-1,CDF-rev,N=2-big} and Refs. therein).

The holomorphic sections $\mathbf{V}$ defined in (\ref{V-bold}) belong to
the holomorphic (chiral) ring over the K\"{a}hler-Hodge bundle defined over
the vector multiplets' scalar manifold. Under a finite symplectic
transformation $\mathcal{S}\in $ $Sp\left( 2n_{V}+2,\mathbb{R}\right) $
defined by (\ref{S-call})-(\ref{sympl-cond-2}), $\mathbf{V}$ transform as
\begin{equation}
\mathbf{V}\left( z\right) \overset{\mathcal{S}}{\longrightarrow }\mathcal{S}%
\mathbf{V}^{\prime }\left( z\right) =\exp \left[ -f\left( z^{\prime }\right) %
\right] \mathcal{S}\mathbf{V}^{\prime }\left( z^{\prime }\right) .  \label{1}
\end{equation}
``$z$'' and ``$z^{\prime }$'' collectively denote the scalar field
parametrization (namely, the coordinate frame) before and after the
application of $\mathcal{S}$. Thus, the action of $\mathcal{S}$ generally
induces a (generally non-linear) coordinate transformation
\begin{equation}
z\longrightarrow z^{\prime }.  \label{coord-transf-1}
\end{equation}
Thus, the holomorphic superpotential $W\equiv \left\langle \mathcal{Q},%
\mathbf{V}\left( z\right) \right\rangle \equiv \mathcal{Q}^{T}\Omega \mathbf{%
V}\left( z\right) $ transforms as (recall (\ref{finite-sympl-cond}))
\begin{equation}
W\overset{\mathcal{S}}{\longrightarrow }\exp \left[ -f\left( z^{\prime
}\right) \right] \left\langle \mathcal{Q}^{\prime },\mathbf{V}^{\prime
}\left( z^{\prime }\right) \right\rangle \equiv \exp \left[ -f\left(
z^{\prime }\right) \right] W^{\prime },  \label{2}
\end{equation}
namely with an holomorphic overall factor $\exp \left[ -f\left( z^{\prime
}\right) \right] $. The holomorphic function $f\left( z^{\prime }\right) $
appearing in (\ref{1}) and (\ref{2}) is the gauge function of the K\"{a}hler
transformation induced by $\mathcal{S}$ on the K\"{a}hler potential $%
\mathcal{K}\left( z,\overline{z}\right) \equiv -\ln \left[ i\left\langle
\overline{\mathbf{V}}\left( \overline{z}\right) ,\mathbf{V}\left( z\right)
\right\rangle \right] $ itself (recall Eq. (\ref{1})):
\begin{equation}
\mathcal{K}\left( z,\overline{z}\right) \overset{\mathcal{S}}{%
\longrightarrow }-\ln \left[ i\left\langle \overline{\mathbf{V}}^{\prime
}\left( \overline{z}^{\prime }\right) ,\mathbf{V}^{\prime }\left( z^{\prime
}\right) \right\rangle \right] +f\left( z^{\prime }\right) +\overline{f}%
\left( \overline{z}^{\prime }\right) \equiv \mathcal{K}^{\prime }\left(
z^{\prime },\overline{z}^{\prime }\right) +f\left( z^{\prime }\right) +%
\overline{f}\left( \overline{z}^{\prime }\right) .  \label{3}
\end{equation}

Eqs. (\ref{2}) and (\ref{3}) yield that the covariantly holomorphic
sections\linebreak $\mathcal{V}\left( z,\overline{z}\right) \equiv \exp %
\left[ \mathcal{K}\left( z,\overline{z}\right) /2\right] \mathbf{V}\left(
z\right) $, belonging to the K\"{a}hler-Hodge $U\left( 1\right) $ bundle,
transform under $\mathcal{S}$ as follows (recall (\ref{1}) and (\ref{3})):
\begin{equation}
\mathcal{V}\left( z,\overline{z}\right) \overset{\mathcal{S}}{%
\longrightarrow }\exp \left[ -i\text{Im}\left( f\left( z^{\prime }\right)
\right) \right] \mathcal{SV}^{\prime }\left( z^{\prime },\overline{z}%
^{\prime }\right) ,  \label{4}
\end{equation}
namely with an overall phase (K\"{a}hler-Hodge $U\left( 1\right) $ factor) $%
\exp \left[ -i\text{Im}\left( f\left( z^{\prime }\right) \right) \right] $.
This in turn implies that the $\mathcal{N}=2$ central charge $Z\left( z,%
\overline{z}\right) \equiv \left\langle \mathcal{Q},\mathcal{V}\left( z,%
\overline{z}\right) \right\rangle $ transforms as
\begin{equation}
Z\left( z,\overline{z}\right) \overset{\mathcal{S}}{\longrightarrow }\exp %
\left[ -i\text{Im}\left( f\left( z^{\prime }\right) \right) \right]
Z^{\prime }\left( z^{\prime },\overline{z}^{\prime }\right) .  \label{5}
\end{equation}

A general consequence of Eqs. (\ref{1})-(\ref{5}) is the following.

Under a transformation $\mathcal{S}\in Sp\left( 2n_{V}+2,\mathbb{R}\right) $%
, $W\left( z\right) $ and $Z\left( z,\overline{z}\right) $ are \textit{%
invariant} \textit{iff} $\mathcal{S}$ \textit{does not induce any change in
the coordinates} of the scalar manifold. By looking at the conditions (\ref
{sympl-cond-1})-(\ref{sympl-cond-2}), it is immediate to realize that $%
\mathcal{O}\in \mathcal{PQ}\left( 2n_{V}+2,\mathbb{R}\right) $ represented
by (\ref{O}) is actually the most general element of $Sp\left( 2n_{V}+2,%
\mathbb{R}\right) $ that does not induce any transformation of coordinates
on the scalar manifold, and thus leaves both $W$ and $Z$ (as well as the
corresponding covariant derivatives $D_{i}W$ and $D_{i}Z$) \textit{invariant}%
.

A direct consequence of this is that the effective BH potential \cite{CFM1}
\begin{equation}
V_{BH}\equiv \left| Z\right| ^{2}+g^{i\overline{j}}\left( D_{i}Z\right)
\overline{D}_{\overline{j}}\overline{Z}  \label{VBH-def}
\end{equation}
is also \textit{invariant} under $\mathcal{PQ}\left( 2n_{V}+2,\mathbb{R}%
\right) $:
\begin{equation}
V_{BH}\left( z,\overline{z};\mathcal{Q}\right) \overset{\mathcal{O}}{%
\longrightarrow }V_{BH}\left( z,\overline{z};\mathcal{Q}\right) .
\end{equation}
For this reason, while $\mathcal{PQ}\left( 2n_{V}+2,\mathbb{R}\right) $ can
be efficiently used to investigate the effects of $\frak{F}$ given by (\ref
{F-sympl}) on the attractor points of $V_{BH}$ itself and on the BH entropy
(through the study of the transformation properties of the quartic $G_{4}$%
-invariant $\mathcal{I}_{4}$; see Sec. \ref{Transformation-I4}), its use in
relation to $Z$, $D_{i}Z$ and $V_{BH}$ has some \textit{caveats}, pointed
out at the start of Sec. \ref{VBH-Transf}. The analysis of the latter Sec.
relies on the results of \cite{CFM1} (see also \cite{Kallosh-rev} for a
review, and Refs. therein) on the axion-free supporting charge
configurations, and related supersymmetry properties, in $d$-SK
geometries.\bigskip

We are now going to show that
\begin{equation}
\frak{pq}\left( 2n,\mathbb{R}\right) \subsetneq \frac{\frak{sp}\left(
2n_{V}+2,\mathbb{R}\right) }{\frak{g}_{4}},  \label{res-alg-1}
\end{equation}
which thus implies, through exponential map:
\begin{equation}
\mathcal{PQ}\left( 2n,\mathbb{R}\right) \subsetneq \frac{Sp\left( 2n_{V}+2,%
\mathbb{R}\right) }{G_{4}}.  \label{res-group-1}
\end{equation}
Namely, the PQ symplectic transformations lie in $Sp\left( 2n_{V}+2,\mathbb{R%
}\right) $ outside of the $d=4$ $U$-duality group $G_{4}$, whose Lie algebra
is denoted by $\frak{g}_{4}$ throughout. Thus, (\ref{rel-2}) and (\ref{rel-1}%
) can respectively be recast as
\begin{gather}
\frak{pq}\left( 2n_{V}+2,\mathbb{R}\right) \equiv \frac{\frak{sp}\left(
2n_{V}+2,\mathbb{R}\right) }{\frak{g}_{4}}\cap \frak{lut}\left( 2n_{V}+2,%
\mathbb{R}\right) ;  \notag \\
\Downarrow \exp  \notag \\
\mathcal{PQ}\left( 2n_{V}+2,\mathbb{R}\right) \equiv \frac{Sp\left( 2n_{V}+2,%
\mathbb{R}\right) }{G_{4}}\cap LUT\left( 2n_{V}+2,\mathbb{R}\right) ,
\label{rel-3}
\end{gather}
where ``$\exp $'' denotes the exponential map.

Clearly, (\ref{res-alg-1})-(\ref{rel-3}) hold whenever $\frak{g}_{4}$ is
well defined, for instance in the $\mathcal{N}=2$ models whose vector
multiplets' scalar manifold is a symmetric coset $G_{4}/H_{4}$, with $H_{4}$
being the maximal compact subgroup (with symmetric embedding) of $G_{4}$
itself (see \textit{e.g.} \cite{dWVVP} and Refs. therein; see also \cite
{LA08-Proc} for a recent survey). Besides the \textit{minimally coupled}
\cite{Luciani} $\mathbb{CP}^{n}$ sequence with quadratic prepotential, these
models are given by all symmetric $d$-SK geometries, whose prepotential is
given by (\ref{d-SKG}), with $d_{ijk}$ satisfying the identity \cite
{GST2,CVP}
\begin{equation}
d_{r(pq}d_{ij)k}d^{rkl}=\frac{4}{3}\delta _{(p}^{l}d_{qij)},  \label{id-symm}
\end{equation}
which implies that $d_{ijk}$ and its contravariant counterpart $d^{ijk}$ are
both $G_{5}$-invariant (scalar-independent) tensors (see Sec. \ref
{Alternative-I4} for further elucidation). Moreover, for all $d$-SK
geometries a ``minimal'' $G_{4}\equiv SO\left( 1,1\right) \times _{s}\mathbb{%
R}^{n_{V}}$ always exists (see Footnote 3).

Furthermore, for a symmetric $d$-SK geometry, the expression of the unique
quartic invariant polynomial $\mathcal{I}_{4}\left( \mathcal{Q}\right) $ of
the symplectic repr. $\mathbf{R}$ of $G_{4}$ reads (in the ``special
coordinates'' sympletic basis \cite{FG1}):
\begin{equation}
\mathcal{I}_{4}\left( \mathcal{Q}\right) \equiv -\left( p^{0}\right)
^{2}q_{0}^{2}-\left( p^{i}q_{i}\right) ^{2}-2p^{0}q_{0}p^{i}q_{i}+4q_{0}%
\mathcal{I}_{3}\left( p\right) -4p^{0}\mathcal{I}_{3}\left( q\right)
+4\left\{ \mathcal{I}_{3}\left( p\right) ,\mathcal{I}_{3}\left( q\right)
\right\} ,  \label{I4-bare-1}
\end{equation}
where
\begin{equation}
\mathcal{I}_{3}\left( p\right) \equiv \frac{1}{3!}d_{ijk}p^{i}p^{j}p^{k};~%
\mathcal{I}_{3}\left( q\right) \equiv \frac{1}{3!}d^{ijk}q_{i}q_{j}q_{k};~%
\left\{ \mathcal{I}_{3}\left( p\right) ,\mathcal{I}_{3}\left( q\right)
\right\} \equiv \frac{\partial \mathcal{I}_{3}\left( p\right) }{\partial
p^{i}}\frac{\partial \mathcal{I}_{3}\left( q\right) }{\partial q_{i}}.
\label{defs-I4-bare}
\end{equation}

In $d$-SK geometries, the manifestly $\left( \frak{g}_{5}\oplus \frak{so}%
\left( 1,1\right) \right) $-covariant form of the symplectic embedding of
the infinitesimal transformation of the $G_{4}$ is provided by the following
$2\left( n_{V}+1\right) \times 2\left( n_{V}+1\right) $ matrix ($%
i,j,k=1,...,n_{V}$) \cite{ADFT-flat-1}:
\begin{equation}
\frak{X}\equiv \left(
\begin{array}{cccc}
3\lambda & b_{j} & 0 & 0^{j} \\
c^{i} & \mathcal{A}_{j}^{i}+\lambda \delta _{j}^{i} & 0^{i} & d^{ijk}b_{k}
\\
0 & 0_{j} & -3\lambda & -c^{j} \\
0_{i} & d_{ijk}c^{k} & -b_{i} & \mathcal{A}_{i}^{j}-\lambda \delta _{i}^{j}
\end{array}
\right) ,  \label{inf-1}
\end{equation}
where $\mathcal{A}_{j}^{i}$ is the electric-magnetic representation of the $%
\frak{g}_{5}$ algebra, $\lambda $ is the $\frak{so}\left( 1,1\right) $
parameter, $c^{i}$ are the parameters of the PQ axion-shift transformations $%
\frak{l}_{+2}$, and $b_{i}$ are the parameters of the additional
transformations $\frak{l}_{-2}^{\prime }$, not implementable on the vector
potentials $A^{0}$, $A^{i}$, which complete the algebra to $\frak{g}_{4}$
(subscripts denote weights w.r.t. $\frak{so}\left( 1,1\right) $):
\begin{equation}
\frak{g}_{4}=\left( \frak{g}_{5}\right) _{0}\oplus \left( \frak{so}\left(
1,1\right) \right) _{0}\oplus \frak{l}_{+2}\oplus \frak{l}_{-2}.
\end{equation}

Thus, the matrix $\frak{X}$ given by (\ref{inf-1}) realizes the Lie algebra $%
\frak{g}_{4}$ of the $U$-duality group $G_{4}$ in its symplectic irrepr. $%
\mathbf{R}$, defining the embedding (\ref{G-sympl-embed}). By comparing the
matrix $\frak{X}$ given by (\ref{inf-1}) with the infinitesimal form of $%
\mathcal{O}$ given by (\ref{O}), namely with the strictly lower triangular
matrix
\begin{equation}
\mathcal{O}_{inf}=\left(
\begin{array}{cc}
0_{2} & 0_{2} \\
\frak{W}_{2} & 0_{2}
\end{array}
\right) \in \frak{pq}\left( 2n_{V}+2,\mathbb{R}\right) ,
\end{equation}
one can conclude that results (\ref{res-alg-1}), and thus (\ref{res-group-1}%
), hold.

\subsubsection{\label{Stringy}Stringy Origin}

It is here worth briefly commenting on the stringy origin of the components
of the matrix $\mathcal{W}_{\Lambda \Sigma }$ given by (\ref{def2}). For
more details, and a list of Refs., we address the reader \textit{e.g.} to
the treatment of \cite{N=2-Quantum,HKTY,Mohaupt-review}.

In Type $IIA$ compactifications over Calabi-Yau threefolds ($CY_{3}$), it
holds that
\begin{equation}
\mathcal{W}_{0i}\equiv \mathbf{c}_{i}=\frac{c_{2,i}}{24}\equiv \frac{%
c_{2}\cdot J_{i}}{24}=\frac{1}{24}\int_{CY_{3}}c_{2}\wedge J_{i},
\label{2nd-Chern-class}
\end{equation}
where $c_{2}$ is the second Chern class\footnote{%
Note that, \textit{e.g.} in presence of $R^{2}$-corrections, the second
Chern class also contributes non-homogeneously to the BH entropy (see
\textit{e.g.} \cite{MSW,Vafa}).} of $CY_{3}$, and $\left\{ J_{i}\right\}
_{i=1,...,n_{V}}$ is a basis of $H^{2}\left( CY_{3},\mathbb{R}\right) $, the
second cohomology group of $CY_{3}$.

Moreover, the coefficients of $\mathbf{F}$ (as given by Eq. (\ref{F-bold}))
have the following stringy interpretation \cite{N=2-Quantum, Alvarez-Gaume,
Grisaru,CDLOGP}:
\begin{eqnarray}
\frac{1}{3!}\text{R}ed_{ijk} &=&\mathbf{C}_{ijk};  \label{tue-1} \\
\frac{1}{3!}\text{Im}d_{000} &=&-\frac{\zeta \left( 3\right) }{\left( 2\pi
\right) ^{3}}\chi ,  \label{i-lambda}
\end{eqnarray}
where $\mathbf{C}_{ijk}$ and $\chi \ $respectively are the classical triple
intersection numbers\footnote{%
Actually, quantum (perturbative and non-perturbative) effects can also
affect R$ed_{ijk}$, \textit{i.e.} (through Eq. (\ref{tue-1})) the classical
triple intersection numbers (see \textit{e.g.} \cite{N=2-Quantum, HKTY}, and
Refs. therein).} and Euler character of the $CY_{3}$, and $\zeta $ is the
Riemann zeta function.

Notice that the other components of $\mathcal{W}_{\Lambda \Sigma }$, namely $%
\mathcal{W}_{00}\equiv \varrho $ and $\mathcal{W}_{ij}\equiv \Theta _{ij}$,
do not have an interpretation in terms of topological invariants of the
internal manifold (see \textit{e.g.} the discussion in \cite{HKTY}), \textit{%
at least} in the compactification framework under consideration. For this
reason, they are usually disregarded in the stringy literature (see \textit{%
e.g.} \cite{N=2-Quantum}, in particular the discussion of Eq. (3.48)
therein; see also \cite{Shmakova}). However, it is worth pointing out that $%
\mathcal{W}_{00}$ and $\mathcal{W}_{ij}$ are important for fixing the
integral basis for $\mathbf{V}$ itself (see \textit{e.g.} the discussion in
\cite{13-of-HKTY-2,14-of-HKTY-2,HKTY}.

When setting $\varrho =\Theta _{ij}=0$, the transformation (\ref{PQ-charge})
yields
\begin{equation}
\left(
\begin{array}{c}
p^{0} \\
p^{i} \\
q_{0} \\
q_{i}
\end{array}
\right) \overset{\mathcal{O}^{-1}}{\longrightarrow }\left(
\begin{array}{l}
p^{\prime 0} \\
p^{\prime i} \\
q_{0}^{\prime }-\mathbf{c}_{j}p^{\prime j} \\
q_{i}^{\prime }-\mathbf{c}_{i}p^{\prime 0}
\end{array}
\right) ,
\end{equation}
which is a Witten theta-shift \cite{Witten-1} of electric charges \textit{via%
} magnetic charges (in a generally axionful background).

Nevertheless, $\mathcal{W}_{00}$ and $\mathcal{W}_{ij}$ are perfectly
consistent in a fully general supergravity analysis, and we will consider
them non-vanishing throughout the applicative developments treated
below.\medskip

In general, the term determined by Re$d_{ijk}$ in the general cubic
prepotential (given by Eqs. (\ref{F-split})-(\ref{F-sympl})) is the \textit{%
leading} one for large values of the scalar fields (\textit{moduli}), and it
defines the purely cubic prepotential (\ref{d-SKG}) of the $d$-SK geometry
of the complex structure (or K\"{a}hler structure) deformation moduli space
of the large volume limit of the internal manifold $CY_{3}$ (in Type II
compactifications). All other terms in Eqs. (\ref{F-split})-(\ref{F-sympl})
define \textit{sub-leading} contributions, which are of quantum perturbative
nature, and consistent with the continuous PQ axion-shift symmetry (\ref
{axion-shift}). All such sub-leading terms, but the purely imaginary
constant determined by $i$Im$d_{000}$ (and eventual renormalization of
classical triple intersection numbers; see Footnote 6), can be taken into
account by means of the group $\mathcal{PQ}\left( 2n_{V}+2,\mathbb{R}\right)
$.

Non-perturbative effects (which can generally traced back to world-sheet
instantons, \textit{i.e.} to non-perturbative phenomena in the non-linear
sigma model) usually exhibit exponential dependence on the moduli, and they
are thus exponentially suppressed in the large volume limit (see \textit{e.g.%
} \cite{N=2-Quantum} and \cite{BM,CC}). They break down the perturbative
continuous PQ axion-shift symmetry (\ref{axion-shift}) to its discrete form,
namely \cite{HKTY}
\begin{equation}
X^{i}\rightarrow X^{i}+1.  \label{axion-shift-discrete}
\end{equation}
In some stringy framework, exponential terms (\textit{e.g.} polylogarithmic
functions) can arise also from quantum perturbative corrections (see \textit{%
e.g.} the discussion in \cite{N=2-Quantum} and \cite{BM,CC}). The effect of
non-perturbative, exponential corrections to cubic prepotentials on the
spectrum and the stability of extremal BH attractors has been recently
addressed in \cite{DT-1}, whose findings confirm the general belief that
non-perturbative correction lift the ``flat'' directions (\textit{if any})
of the perturbative theory\footnote{%
Actually, also quantum perturbative corrections, such as the one given \ by
the term $i$Im$d_{000}$ in (\ref{F-bold}) (with stringy origin given by (\ref
{i-lambda})) can lift (some of the) \textit{``flat directions''} of extremal
BH attractor solutions \cite{BFMS-1}.}. At the level of the prepotential,
this can be traced back to the fact that exponential corrections to the
purely cubic holomorphic prepotential (\ref{d-SKG}) $d$-SK geometries (of
the kind given by Eq. (4.1) of \cite{DT-1}) affect the geometry properties
of the scalar manifold itself.

\subsection{\label{Applications}Application to Black Hole Attractors,
Entropy and Supersymmetry}

As pointed out in Sec. \ref{Rel-U-Duality}, the Peccei-Quinn symplectic
group $\mathcal{PQ}\left( 2n_{V}+2,\mathbb{R}\right) $ is a proper subgroup
of $\frac{Sp\left( 2n_{V}+2,\mathbb{R}\right) }{G_{4}}$. The latter is the
most general group acting linearly on the charges $\mathcal{Q}$ which can
change the value and possibly the sign of the unique quartic invariant $%
\mathcal{I}_{4}\left( \mathcal{Q}\right) $ of the symplectic (ir)repr. $%
\mathbf{R}$ of $G_{4}$ itself.

In the following treatment, within the manifestly $G_{5}$-covariant
``special coordinates'' symplectic frame, we will analyse how $\mathcal{PQ}%
\left( 2n_{V}+2,\mathbb{R}\right) $ acts on $\mathcal{I}_{4}\left( \mathcal{Q%
}\right) $, on the non-degenerate critical points of the effective BH
potential $V_{BH}$ (\textit{alias} extremal BH attractors) \cite{AM-Refs},
and on their supersymmetry properties. We will work within the $d$-SK
geometries determined by the prepotential (\ref{d-SKG}). When they involve
the contravariant tensor $d^{ijk}$, the results on the transformation
properties of $\mathcal{I}_{4}$ generally hold only for $d$-SK geometries
such that the coset $G_{4}/H_{4}$ is \textit{symmetric} (see \textit{e.g.}
\cite{dWVVP}, and Refs. therein).

By suitably adapting its action, $\mathcal{PQ}\left( 2n_{V}+2,\mathbb{R}%
\right) $ reveals to be a very effective tool to investigate the effect of
the quantum perturbative sub-leading corrections (\ref{F-sympl}) to the
leading $d$-SK prepotential (\ref{d-SKG}), some of which have a topological
interpretation (see Sec. \ref{Stringy}).

We anticipate that, under certain conditions on the ratio between the
charges $\mathcal{Q}$ and the parameters $\left( \varrho ,\mathbf{c}%
_{i},\Theta _{ij}\right) $ of the finite PQ transformation $\mathcal{O}$
(given by Eq. (\ref{O}) and (\ref{def2})), the action of $\mathcal{PQ}\left(
2n_{V}+2,\mathbb{R}\right) $ can give rise to a ``transition'' among the
various orbits of $\mathbf{R}$ of $G_{4}$, which in turn changes the
supersymmetry-preserving features of the extremal BH attractor solutions%
\footnote{%
Thus, our results should have interesting connections with the $d=3$
timelike-reduced geodesic formalism and results of \cite{Trig-d=3}, whose
thorough investigation we leave for further future study. For some
developments in a $d=4$ framework, see \cite{K-can} (and also \cite{ADFT-rev}%
).}.

\subsubsection{\label{Transformation-I4}Transformation of $\mathcal{I}_{4}$}

We start and apply the finite transformation\footnote{%
We consider $\mathcal{O}^{-1}$ rather than $\mathcal{O}$ (a choice which is
clearly immaterial at group level) because operationally (as discussed in
\cite{N=2-Quantum}) one would like to include the effects of the sub-leading
$\left( \varrho ,\mathbf{c}_{i},\Theta _{ij}\right) $-dependent terms in the
prepotential (\ref{F-split})-(\ref{F-sympl}) on the Bekenstein-Hawking BH
entropy \cite{BH1} by simply performing the computations within the purely
cubic prepotential (\ref{d-SKG}) (see \textit{e.g.} the analysis of \cite
{CFM1}) and then by applying the transformation $\mathcal{O}^{-1}$ on $%
\mathcal{Q}$. Note that we will not deal here with the term $\frac{i}{3!}$Im$%
d_{000}\left( X^{0}\right) ^{2}$ in (\ref{F-bold}), which has been recently
studied in \cite{Raju-1}.} $\mathcal{O}^{-1}\in \mathcal{PQ}\left( 2n_{V}+2,%
\mathbb{R}\right) $ (given by (\ref{PQ-charge})) to the $G_{4}$-invariant
quartic polynomial $\mathcal{I}_{4}\left( \mathcal{Q}\right) $ given by (\ref
{I4-bare-1})-(\ref{defs-I4-bare}). Thus, after some algebra, the following
result is achieved:
\begin{equation}
\mathcal{PQ}\left( 2n_{V}+2,\mathbb{R}\right) \ni \mathcal{O}^{-1}:\mathcal{I%
}_{4}\left( \mathcal{Q}\right) \longrightarrow \mathcal{I}_{4}^{\prime
}\left( \mathcal{O}^{-1}\mathcal{Q}^{\prime }\right) =\mathcal{I}_{4}\left(
\mathcal{O}^{-1}\mathcal{Q}^{\prime }\right) =\mathcal{I}_{4}\left( \mathcal{%
Q}^{\prime }\right) +\mathbf{I}_{4}\left( \mathcal{Q}^{\prime };\varrho ,%
\mathbf{c}_{i},\Theta _{ij}\right) ,  \label{PQ-I4}
\end{equation}
where the quartic quantity $\mathbf{I}_{4}$, describing the
``PQ-deformation'' of $\mathcal{I}_{4}\left( \mathcal{Q}\right) $, is given
by the following expression\footnote{%
Throughout the subsequent treatment, we omit the priming of the $\mathcal{O}%
^{-1}$-transformed charges.}:
\begin{eqnarray}
\mathbf{I}_{4}\left( \mathcal{Q};\varrho ,\mathbf{c}_{i},\Theta _{ij}\right)
&\equiv &2\left( p^{0}\right) ^{4}\left( \frac{1}{3}d^{ijk}\mathbf{c}_{i}%
\mathbf{c}_{j}\mathbf{c}_{k}-\frac{1}{2}\varrho ^{2}\right)  \notag \\
&&  \notag \\
&&+2\left( p^{0}\right) ^{3}\left( \varrho q_{0}-\varrho \mathbf{c}%
_{i}p^{i}-d^{ijk}q_{i}\mathbf{c}_{j}\mathbf{c}_{k}+d^{ijk}\mathbf{c}_{i}%
\mathbf{c}_{j}\Theta _{kl}p^{l}\right)  \notag \\
&&  \notag \\
&&+2\left( p^{0}\right) ^{2}\left(
\begin{array}{l}
-2\left( \mathbf{c}_{i}p^{i}\right) ^{2}+2q_{0}\mathbf{c}_{i}p^{i}+\varrho
p^{i}q_{i}-\varrho \Theta _{ij}p^{i}p^{j}-2d^{ijk}q_{i}\mathbf{c}_{j}\Theta
_{kl}p^{l} \\
\\
+d^{ijk}\mathbf{c}_{i}\Theta _{jl}\Theta _{km}p^{l}p^{m}+\frac{1}{2}%
d_{ijk}d^{ilm}\mathbf{c}_{l}\mathbf{c}_{m}p^{j}p^{k}+d^{ijk}q_{i}q_{j}%
\mathbf{c}_{k}
\end{array}
\right)  \notag \\
&&  \notag \\
&&+2p^{0}\left(
\begin{array}{l}
2p^{i}q_{i}\mathbf{c}_{j}p^{j}-2\mathbf{c}_{i}\Theta
_{jk}p^{i}p^{j}p^{k}+q_{0}\Theta _{ij}p^{i}p^{j}-\frac{1}{3}\varrho
d_{ijk}p^{i}p^{j}p^{k} \\
\\
+d^{ijk}q_{i}q_{j}\Theta _{kl}p^{l}-d^{ijk}q_{i}\Theta _{jl}\Theta
_{km}p^{l}p^{m}+\frac{1}{3}d^{ijk}\Theta _{il}\Theta _{jm}\Theta
_{kn}p^{l}p^{m}p^{n} \\
\\
-d_{ijk}d^{ilm}p^{j}p^{k}q_{l}\mathbf{c}_{m}+d_{ijk}d^{ilm}p^{j}p^{k}\mathbf{%
c}_{l}\Theta _{ms}p^{s}
\end{array}
\right)  \notag \\
&&  \notag \\
&&-\left( \Theta _{ij}p^{i}p^{j}\right) ^{2}+2p^{i}q_{i}\Theta
_{jk}p^{j}p^{k}-\frac{2}{3}\mathbf{c}_{l}p^{l}d_{ijk}p^{i}p^{j}p^{k}  \notag
\\
&&  \notag \\
&&-2d_{ijk}d^{ilm}p^{j}p^{k}q_{l}\Theta
_{ms}p^{s}+d_{ijk}d^{ilm}p^{j}p^{k}\Theta _{ls}\Theta _{mt}p^{s}p^{t}.
\label{PQ-Ibold4}
\end{eqnarray}
Note that the degree-$4$ homogeneity of $\mathcal{I}_{4}$ in the charges is
not spoiled, due to the linearity of the action of $\mathcal{PQ}\left(
2n_{V}+2,\mathbb{R}\right) $ on the charges themselves.

We now analyse various particular (both ``large'' and ``small'') charge
configurations, showing how the action of $\mathcal{PQ}\left( 2n_{V}+2,%
\mathbb{R}\right) $ can give rise to two types of phenomena, both
corresponding to switching among \ different $\mathbb{R}$-orbits:

\begin{itemize}
\item  change of sign of $\mathcal{I}_{4}$:
\begin{equation}
\mathcal{I}_{4}\left( \mathcal{Q}\right) \gtrless 0\overset{\mathcal{PQ}}{%
\longrightarrow }\mathcal{I}_{4}\left( \mathcal{Q}\right) +\mathbf{I}%
_{4}\left( \mathcal{Q};\varrho ,\mathbf{c}_{i},\Theta _{ij}\right) \lessgtr
0,  \label{1-phen}
\end{equation}
corresponding to a switch between different ``large'' $\mathbb{R}$-orbits
\cite{BFGM1};

\item  generation of a non-vanishing $\mathcal{I}_{4}$:
\begin{equation}
\mathcal{I}_{4}\left( \mathcal{Q}\right) =0\overset{\mathcal{PQ}}{%
\longrightarrow }\mathcal{I}_{4}\left( \mathcal{Q}\right) +\mathbf{I}%
_{4}\left( \mathcal{Q};\varrho ,\mathbf{c}_{i},\Theta _{ij}\right) \gtrless
0,  \label{2.1-phen}
\end{equation}
or the other way around, generation of a vanishing $\mathcal{I}_{4}$:
\begin{equation}
\mathcal{I}_{4}\left( \mathcal{Q}\right) \gtrless 0\overset{\mathcal{PQ}}{%
\longrightarrow }\mathcal{I}_{4}\left( \mathcal{Q}\right) +\mathbf{I}%
_{4}\left( \mathcal{Q};\varrho ,\mathbf{c}_{i},\Theta _{ij}\right) =0,
\label{2.2-phen}
\end{equation}
both corresponding to a switch between a ``large'' and a ``small'' $\mathbb{R%
} $-orbit (usually named ``charge orbit'').\medskip
\end{itemize}

Some comments on the meaning of Eqs. (\ref{1-phen})-(\ref{2.2-phen}) are in
order.

\begin{itemize}
\item  Firstly, let us recall that, through the Bekenstein-Hawking formula (%
\ref{Bek}), ``large'' and ``small'' charge orbits respectively corresponds
to $\mathcal{I}_{4}\neq 0$ and $\mathcal{I}_{4}=0$; furthermore, ``small''
orbits split in lightlike ($3$-charge), critical ($2$-charge) and
doubly-critical ($1$-charge) ones \cite{FG1,CFMZ1,ADFT-FO-1,ICL-1,CFM2}.
\end{itemize}

Then, the general treatment of Sec. \ref{GT} implies that, in presence of $%
\left( \varrho ,\mathbf{c}_{i},\Theta _{ij}\right) $-dependent sub-leading
contributions (\ref{F-sympl}) (recall the change of notation (\ref{def2}))
to the purely cubic prepotential (\ref{d-SKG}) of $d$-SK geometry, the BH
entropy $S$ becomes $\left( \varrho ,\mathbf{c}_{i},\Theta _{ij}\right) $%
-dependent:
\begin{equation}
\frac{S}{\pi }=\frac{A}{4}=\sqrt{\left| \mathcal{I}_{4}\left( \mathcal{Q}%
\right) +\mathbf{I}_{4}\left( \mathcal{Q};\varrho ,\mathbf{c}_{i},\Theta
_{ij}\right) \right| },  \label{Bek-1}
\end{equation}
where $\mathbf{I}_{4}\left( \mathcal{Q};\varrho ,\mathbf{c}_{i},\Theta
_{ij}\right) $ is defined in (\ref{PQ-Ibold4}). Consequently, depending on
the relations between $\mathcal{I}_{4}\left( \mathcal{Q}\right) $ and $%
\mathbf{I}_{4}\left( \mathcal{Q};\varrho ,\mathbf{c}_{i},\Theta _{ij}\right)
$, the phenomena (\ref{1-phen})-(\ref{2.2-phen}) can occur, and the ones
related to $\mathbf{c}_{i}$ have, by virtue of (\ref{2nd-Chern-class}), a
clear topological interpretation within Type $II$ $CY_{3}$-compactifications.

It should be remarked that the geometry of the \textit{symmetric} coset $%
G_{4}/H_{4}$ is unaffected by the action of $Sp\left( 2n_{V}+2,\mathbb{R}%
\right) $ (which just produces a change of coordinates; see Sec. \ref
{Rel-U-Duality}), and thus \textit{a fortiori} by the action of its proper
subgroup $\mathcal{PQ}\left( 2n_{V}+2,\mathbb{R}\right) $. Furthermore, by
virtue of the treatment of Sec. \ref{Rel-U-Duality}, $\mathcal{PQ}\left(
2n_{V}+2,\mathbb{R}\right) $ does not act on the coordinates of the scalar
manifolds, and thus does not induce any K\"{a}hler gauge transformation (\ref
{3}) on $\mathcal{K}$, nor any holomorphic scaling (\ref{2}) on $W$ (and $%
D_{i}W$) and local phase transformation (\ref{5}) on $Z$ (and $D_{i}Z$)
itself. Thus, the only effect of $\mathcal{PQ}\left( 2n_{V}+2,\mathbb{R}%
\right) $ on the BH effective potential $V_{BH}$ and its non-degenerate
critical points (\textit{alias} extremal BH attractors) \cite{AM-Refs} is a $%
\left( \varrho ,\mathbf{c}_{i},\Theta _{ij}\right) $-dependent
transformation of the charge vector $\mathcal{Q}$, as given by (\ref
{PQ-charge}). This fact will allow us to analyse the axion-free-supporting
nature of the BH charge configurations in presence of non-vanishing
parameters $\varrho $, $\mathbf{c}_{i}$ and $\Theta _{ij}$ by relying on the
results of \cite{CFM1} (holding for generic (\ref{d-SKG})). The results
recently obtained in\ Sec. 3 of \cite{DT-1} are an expected confirmation of
all this reasoning.

By virtue of the transition from (\ref{Bek}) to (\ref{Bek-1}) through (\ref
{PQ-I4}), $Sp\left( 2n_{V}+2,\mathbb{R}\right) $ (and therefore its proper
subgroup $\mathcal{PQ}\left( 2n_{V}+2,\mathbb{R}\right) $) does \textit{not}
affect the geometry of the scalar manifold, but it may affect the
``magnitude'' of the near-horizon space-time BH background, since its action
may change the event horizon area $A$ of the extremal BH, and thus the
(semi)classical Bekenstein-Hawking BH entropy $S$. The phenomena described
by Eqs. (\ref{1-phen})-(\ref{2.2-phen}) correspond to $\left( \varrho ,%
\mathbf{c}_{i},\Theta _{ij}\right) $-dependent transformations moving from
one charge orbit to another in the representation space $\mathbf{R}$ of $%
G_{4}$.

The geometry and the classification of BH charge orbits (and related \textit{%
``moduli spaces''}\footnote{%
This has been recently confirmed by the analysis of the particular model of
Sec. 3 of \cite{DT-1}.}) is not affected by $Sp\left( 2n_{V}+2,\mathbb{R}%
\right) $ (and therefore by $\mathcal{PQ}\left( 2n_{V}+2,\mathbb{R}\right) $%
), but symplectic transformations can induce ``transmutations'' of the
nature of the charge vector $\mathcal{Q}\longrightarrow \mathcal{Q}^{\left(
\prime \right) }\left( \varrho ,\mathbf{c}_{i},\Theta _{ij}\right) $, and
thus of its supersymmetry preserving properties. As we will see in the case
study considered in Sec. \ref{Analysis-Configs}, in the case of $\mathcal{PQ}%
\left( 2n_{V}+2,\mathbb{R}\right) $ the actual occurrence of these phenomena
depends on the very relations between $\mathcal{Q}$ and the transformtaion
parameters $\left( \varrho ,\mathbf{c}_{i},\Theta _{ij}\right) $ themselves.

\subsubsection{\label{Analysis-Configs}Analysis of ``Large'' and ``Small''
Configurations}

The above treatment will be further clarified by the various examples which
we are going to treat, generalising and systematically developing some
points mentioned in \cite{N=2-Quantum}. We will make extensive use of
formul\ae\ (\ref{PQ-charge}) and (\ref{PQ-I4})-(\ref{Bek-1}).

\begin{enumerate}
\item  \textbf{``Large'' }$\left( p^{0},q_{0}\right) $ \textbf{(Kaluza-Klein)%
} \textbf{configuration}. It supports non-BPS $Z_{H}\neq 0$ (possibly
axion-free \cite{CFM1}) attractors, and it is the supergravity analogue of $%
D0$-$D6$ configuration in Type $II$:
\begin{equation}
\mathcal{Q}\equiv \left( p^{0},0,q_{0},0\right) ^{T}\Rightarrow \mathcal{I}%
_{4}\left( \mathcal{Q}\right) =-\left( p^{0}\right) ^{2}q_{0}^{2}<0.
\end{equation}
The action of $\mathcal{PQ}\left( 2n_{V}+2,\mathbb{R}\right) $ reads
\begin{equation}
\left(
\begin{array}{l}
p^{0} \\
0 \\
q_{0} \\
0
\end{array}
\right) \overset{\mathcal{O}^{-1}}{\longrightarrow }\left(
\begin{array}{l}
p^{0} \\
0 \\
q_{0}-\varrho p^{0} \\
-\mathbf{c}_{i}p^{0}
\end{array}
\right) ,
\end{equation}
and thus it generates $\mathbf{c}_{i}$-dependent electric charges $q_{i}$'s,
which in Type $II$ compactifications corresponds to a stack of $D2$ branes
depending on the components of the second Chern class $c_{2}$ of $CY_{3}$
(recall Eq. (\ref{2nd-Chern-class})). The corresponding transformation of $%
\mathcal{I}_{4}$ reads
\begin{equation}
-\left( p^{0}\right) ^{2}q_{0}^{2}<0\overset{\mathcal{O}^{-1}}{%
\longrightarrow }\left( p^{0}\right) ^{4}\left[ \frac{2}{3}d^{ijk}\mathbf{c}%
_{i}\mathbf{c}_{j}\mathbf{c}_{k}-\left( \frac{q_{0}}{p^{0}}-\varrho \right)
^{2}\right] \gtreqless 0.  \label{pre-rell}
\end{equation}
Thus, depending on whether
\begin{equation}
\frac{2}{3}d^{ijk}\mathbf{c}_{i}\mathbf{c}_{j}\mathbf{c}_{k}\gtreqless
\left( \frac{q_{0}}{p^{0}}-\varrho \right) ^{2},  \label{rell}
\end{equation}
a ``large'' ($\mathcal{I}_{4}>0$:BPS or non-BPS $Z_{H}=0$), a ``small'' ($%
\mathcal{I}_{4}=0$:BPS or non-BPS), or a ``large'' non-BPS $Z_{H}\neq 0$ ($%
\mathcal{I}_{4}<0$) BH charge configuration is generated by the action of $%
\mathcal{PQ}\left( 2n_{V}+2,\mathbb{R}\right) $. As anticipated in the above
treatment, (\ref{rell}) shows that the relations among the components of $%
\mathcal{Q}$ and the parameters of the PQ symplectic transformation turn out
to be crucial for the properties of the resulting charge configuration. The
change of the axion-free-supporting nature of this configuration will be
analysed in Sec. \ref{VBH-Transf}.

\item  \textbf{``Large'' }$\left( p^{0},q_{i}\right) $ \textbf{(``electric'')%
} \textbf{configuration}. Depending on $\mathcal{I}_{4}\left( \mathcal{Q}%
\right) \gtrless 0$, it supports all kind of attractors (possibly axion-free
\cite{CFM1}). It is the supergravity analogue of $D2$-$D6$ configuration in
Type $II$:
\begin{equation}
\mathcal{Q}\equiv \left( p^{0},0,0,q_{i}\right) ^{T}\Rightarrow \mathcal{I}%
_{4}\left( \mathcal{Q}\right) =-\frac{2}{3}p^{0}d^{ijk}q_{i}q_{j}q_{k}%
\gtrless 0.
\end{equation}
The action of $\mathcal{PQ}\left( 2n_{V}+2,\mathbb{R}\right) $ is
\begin{equation}
\left(
\begin{array}{l}
p^{0} \\
0 \\
0 \\
q_{i}
\end{array}
\right) \overset{\mathcal{O}^{-1}}{\longrightarrow }\left(
\begin{array}{l}
p^{0} \\
0 \\
-\varrho p^{0} \\
q_{i}-\mathbf{c}_{i}p^{0}
\end{array}
\right) ,
\end{equation}
and thus it generates a $\varrho $-dependent electric charge $q_{0}$. The
corresponding transformation of $\mathcal{I}_{4}$ reads
\begin{gather}
-\frac{2}{3}p^{0}d^{ijk}q_{i}q_{j}q_{k}\gtrless 0  \notag \\
\downarrow \mathcal{O}^{-1}  \notag \\
-\frac{2}{3}p^{0}d^{ijk}q_{i}q_{j}q_{k}+2\left( p^{0}\right) ^{2}\left[
\left( \frac{1}{3}d^{ijk}\mathbf{c}_{i}\mathbf{c}_{j}\mathbf{c}_{k}-\frac{1}{%
2}\varrho ^{2}\right) \left( p^{0}\right) ^{2}-p^{0}d^{ijk}q_{i}\mathbf{c}%
_{j}\mathbf{c}_{k}+d^{ijk}q_{i}q_{j}\mathbf{c}_{k}\right] \gtreqless 0.
\label{rell-2}
\end{gather}
Thus, depending on the sign (or on the vanishing) of the quantity in the
last line of (\ref{rell-2}), the same comments made for configuration
\textbf{1} hold in this case. The change of the axion-free-supporting nature
of this configuration will be analysed in Sec. \ref{VBH-Transf}.

\item  \textbf{``Large'' }$\left( p^{i},q_{0}\right) $ \textbf{(``magnetic'')%
} \textbf{configuration}. It is the \textit{``electric-magnetic dual''} of
the ``electric'' configuration \textbf{2}. It is then interesting to compare
the action of $\mathcal{PQ}\left( 2n_{V}+2,\mathbb{R}\right) $ (which is
asymmetric on magnetic and electric charges) on configurations \textbf{2}
and \textbf{3}. Depending on $\mathcal{I}_{4}\left( \mathcal{Q}\right)
\gtrless 0$, this configuration supports all kind of attractors (possibly
axion-free \cite{CFM1}). It is the supergravity analogue of $D0$-$D4$
configuration in Type $II$:
\begin{equation}
\mathcal{Q}\equiv \left( 0,p^{i},q_{0},0\right) ^{T}\Rightarrow \mathcal{I}%
_{4}\left( \mathcal{Q}\right) =\frac{2}{3}q_{0}d_{ijk}p^{i}p^{j}p^{k}%
\gtrless 0.
\end{equation}
The action of $\mathcal{PQ}\left( 2n_{V}+2,\mathbb{R}\right) $ is
\begin{equation}
\left(
\begin{array}{l}
0 \\
p^{i} \\
q_{0} \\
0
\end{array}
\right) \overset{\mathcal{O}^{-1}}{\longrightarrow }\left(
\begin{array}{l}
0 \\
p^{i} \\
q_{0}-\mathbf{c}_{j}p^{j} \\
-\Theta _{ij}p^{j}
\end{array}
\right) ,
\end{equation}
and thus it generates $\Theta _{ij}$-dependent electric charges $q_{i}$'s.
The corresponding transformation of $\mathcal{I}_{4}$ reads
\begin{gather}
\frac{2}{3}q_{0}d_{ijk}p^{i}p^{j}p^{k}\gtrless 0  \notag \\
\downarrow \mathcal{O}^{-1}  \notag \\
\frac{2}{3}q_{0}d_{ijk}p^{i}p^{j}p^{k}-\left( \Theta _{ij}p^{i}p^{j}\right)
^{2}-\frac{2}{3}\mathbf{c}%
_{l}p^{l}d_{ijk}p^{i}p^{j}p^{k}+d_{ijk}d^{ilm}p^{j}p^{k}\Theta _{ls}\Theta
_{mt}p^{s}p^{t}\gtreqless 0.  \label{rell-3}
\end{gather}
Thus, depending on the sign (or on the vanishing) of the quantity in the
last line of (\ref{rell-3}), the same comments as made for above
configurations hold. The change of the axion-free-supporting nature of this
configuration will be analysed in Sec. \ref{VBH-Transf}. Note that for $%
\Theta _{ij}=0$, an example treated in \cite{N=2-Quantum} is recovered.

\item  \textbf{``Small'' \textit{lightlike} (}$3$\textbf{-charge) }$q_{i}$
\textbf{(``electric'')} \textbf{configuration}. This is the limit $p^{0}=0$
of configuration \textbf{2}. In Type $II$, it corresponds to only $D2$
branes:
\begin{equation}
\mathcal{Q}\equiv \left( 0,0,0,q_{i}\right) ^{T}\Rightarrow \mathcal{I}%
_{4}\left( \mathcal{Q}\right) =0,
\end{equation}
such that (recall definition (\ref{defs-I4-bare}))
\begin{equation}
\mathcal{I}_{3}\left( q\right) \neq 0,
\end{equation}
corresponding to a ``large'' BH in $d=5$, with near-horizon geometry $%
AdS_{2}\times S^{3}$ (see \textit{e.g.} \cite{CFM1}, and Refs. therein).
Since there are no magnetic charges, $\mathcal{PQ}\left( 2n_{V}+2,\mathbb{R}%
\right) $ is inactive on this configuration, which is thus left unchanged:
\begin{equation}
\left(
\begin{array}{l}
0 \\
0 \\
0 \\
q_{i}
\end{array}
\right) \overset{\mathcal{O}^{-1}}{\longrightarrow }\left(
\begin{array}{l}
0 \\
0 \\
0 \\
q_{i}
\end{array}
\right) .  \label{PQ-inactive}
\end{equation}

\item  \textbf{``Small'' \textit{lightlike} (}$3$\textbf{-charge) }$p^{i}$
\textbf{(``magnetic'')} \textbf{configuration}. This is the limit $q_{0}=0$
of configuration \textbf{3}. In Type $II$, it corresponds to only $D4$
branes:
\begin{equation}
\mathcal{Q}\equiv \left( 0,p^{i},0,0\right) ^{T}\Rightarrow \mathcal{I}%
_{4}\left( \mathcal{Q}\right) =0,
\end{equation}
such that (recall definition (\ref{defs-I4-bare}))
\begin{equation}
\mathcal{I}_{3}\left( p\right) \neq 0,
\end{equation}
corresponding to a ``large'' black string in $d=5$, with near-horizon
geometry $AdS_{3}\times S^{2}$ (see \textit{e.g.} \cite{CFM1}, and Refs.
therein). This configuration is the \textit{``electric-magnetic dual''} of
the ``electric'' configuration \textbf{4}. However, differently from what
happens for configuration \textbf{4}, $\mathcal{PQ}\left( 2n_{V}+2,\mathbb{R}%
\right) $ is active in this case (due to its asymmetric action on electric
and magnetic charges):
\begin{equation}
\left(
\begin{array}{l}
0 \\
p^{i} \\
0 \\
0
\end{array}
\right) \overset{\mathcal{O}^{-1}}{\longrightarrow }\left(
\begin{array}{l}
0 \\
p^{i} \\
-\mathbf{c}_{j}p^{j} \\
-\Theta _{ij}p^{j}
\end{array}
\right) .  \label{PQ-active}
\end{equation}
and it generates $\Theta _{ij}$-dependent electric charges $q_{i}$'s, as
well as $\mathbf{c}_{i}$-dependent electric charge $q_{0}$. In Type $II$
compactifications, the latter corresponds to a stack of $D0$ branes
depending on the components of the second Chern class $c_{2}$ of $CY_{3}$
(recall Eq. (\ref{2nd-Chern-class})). The corresponding transformation of $%
\mathcal{I}_{4}$ reads
\begin{equation}
0\overset{\mathcal{O}^{-1}}{\longrightarrow }-\left( \Theta
_{ij}p^{i}p^{j}\right) ^{2}-\frac{2}{3}\mathbf{c}%
_{l}p^{l}d_{ijk}p^{i}p^{j}p^{k}+d_{ijk}d^{ilm}p^{j}p^{k}\Theta _{ls}\Theta
_{mt}p^{s}p^{t}\gtreqless 0.  \label{rell-4}
\end{equation}
Thus, according to (\ref{rell-4}), a ``large'' ($\mathcal{I}_{4}>0$:BPS or
non-BPS $Z_{H}=0$), a ``small'' ($\mathcal{I}_{4}=0$:BPS or non-BPS), or a
``large'' non-BPS $Z_{H}\neq 0$ ($\mathcal{I}_{4}<0$) BH charge
configuration can be generated. In case the quantity in (\ref{rell-4}) does
not vanish, this is an example of phenomenon (\ref{2.1-phen}). Note that for
$\Theta _{ij}=0$, an example treated in \cite{N=2-Quantum} is recovered.

\item  \textbf{``Small'' \textit{critical} (}$2$\textbf{-charge) }$q_{i}$
\textbf{(``electric'')} \textbf{configuration}. This is the limit $\mathcal{I%
}_{3}\left( q\right) =0$ of configuration \textbf{4}. In Type $II$, it
corresponds to only $D2$ branes:
\begin{equation}
\mathcal{Q}\equiv \left( 0,0,0,q_{i}\right) ^{T}\Rightarrow \mathcal{I}%
_{4}\left( \mathcal{Q}\right) =0,
\end{equation}
such that (recall definition (\ref{defs-I4-bare}))
\begin{equation}
\left\{
\begin{array}{l}
\mathcal{I}_{3}\left( q\right) =0; \\
\partial \mathcal{I}_{3}\left( q\right) /\partial q_{i}\neq 0~\text{for~some~%
}i,
\end{array}
\right.
\end{equation}
corresponding to a ``small'' \textit{lightlike} BH in $d=5$. Since there are
no magnetic charges, $\mathcal{PQ}\left( 2n_{V}+2,\mathbb{R}\right) $ is
inactive on this configuration, which is thus left unchanged (see Eq. (\ref
{PQ-inactive})).

\item  \textbf{``Small'' \textit{critical} (}$2$\textbf{-charge) }$p^{i}$
\textbf{(``magnetic'')} \textbf{configuration}. This is the limit $\mathcal{I%
}_{3}\left( q\right) =0$ of configuration \textbf{5}. In Type $II$, it
corresponds to only $D4$ branes:
\begin{equation}
\mathcal{Q}\equiv \left( 0,p^{i},0,0\right) ^{T}\Rightarrow \mathcal{I}%
_{4}\left( \mathcal{Q}\right) =0,
\end{equation}
such that (recall definition (\ref{defs-I4-bare}))
\begin{equation}
\left\{
\begin{array}{l}
\mathcal{I}_{3}\left( p\right) =0; \\
\partial \mathcal{I}_{3}\left( p\right) /\partial p^{i}\neq 0~\text{for~some~%
}i,
\end{array}
\right.
\end{equation}
corresponding to a ``small'' \textit{lightlike} black string in $d=5$. This
configuration is the \textit{``electric-magnetic dual''} of the ``electric''
configuration \textbf{6}. However, differently from what happens for
configuration \textbf{6}, $\mathcal{PQ}\left( 2n_{V}+2,\mathbb{R}\right) $
is active in this case, due to its asymmetric action on electric and
magnetic charges. As given by Eq. (\ref{PQ-active}), $\Theta _{ij}$%
-dependent electric charges $q_{i}$'s and $\mathbf{c}_{i}$-dependent
electric charge $q_{0}$ are generated. The corresponding transformation of $%
\mathcal{I}_{4}$ reads
\begin{equation}
0\overset{\mathcal{O}^{-1}}{\longrightarrow }-\left( \Theta
_{ij}p^{i}p^{j}\right) ^{2}+d_{ijk}d^{ilm}p^{j}p^{k}\Theta _{ls}\Theta
_{mt}p^{s}p^{t}\gtreqless 0.  \label{rell-5}
\end{equation}
Thus, according to (\ref{rell-5}), a ``large'' ($\mathcal{I}_{4}>0$:BPS or
non-BPS $Z_{H}=0$), a ``small'' ($\mathcal{I}_{4}=0$:BPS or non-BPS), or a
``large'' non-BPS $Z_{H}\neq 0$ ($\mathcal{I}_{4}<0$) BH charge
configuration can be generated. In case the quantity in (\ref{rell-5}) does
not vanish, this is an example of phenomenon (\ref{2.1-phen}).

\item  \textbf{``Small'' \textit{doubly-critical} (}$1$\textbf{-charge) }$%
q_{i}$ \textbf{(``electric'')} \textbf{configuration}. This is the limit $%
\partial \mathcal{I}_{3}\left( q\right) /\partial q_{i}=0$ of configuration
\textbf{6}. In Type $II$, it corresponds to only $D2$ branes:
\begin{equation}
\mathcal{Q}\equiv \left( 0,0,0,q_{i}\right) ^{T}\Rightarrow \mathcal{I}%
_{4}\left( \mathcal{Q}\right) =0,
\end{equation}
such that (recall definition (\ref{defs-I4-bare}))
\begin{equation}
\left\{
\begin{array}{l}
\mathcal{I}_{3}\left( q\right) =0; \\
\partial \mathcal{I}_{3}\left( q\right) /\partial q_{i}=0~\forall i;~ \\
q_{i}\neq 0~\text{for~some~}i,
\end{array}
\right.
\end{equation}
corresponding to a ``small'' \textit{critical} BH in $d=5$. Since there are
no magnetic charges, $\mathcal{PQ}\left( 2n_{V}+2,\mathbb{R}\right) $ is
inactive on this configuration, which is thus left unchanged (see Eq. (\ref
{PQ-inactive})).

\item  \textbf{``Small'' \textit{doubly-critical} (}$1$\textbf{-charge) }$%
p^{i}$ \textbf{(``magnetic'')} \textbf{configuration}. This is the limit $%
\partial \mathcal{I}_{3}\left( p\right) /\partial p^{i}=0$ of configuration
\textbf{7}. In Type $II$, it corresponds to only $D4$ branes:
\begin{equation}
\mathcal{Q}\equiv \left( 0,p^{i},0,0\right) ^{T}\Rightarrow \mathcal{I}%
_{4}\left( \mathcal{Q}\right) =0,
\end{equation}
such that (recall definition (\ref{defs-I4-bare}))
\begin{equation}
\left\{
\begin{array}{l}
\mathcal{I}_{3}\left( p\right) =0; \\
\partial \mathcal{I}_{3}\left( p\right) /\partial p^{i}=0~\forall i; \\
p^{i}\neq 0~\text{for~some~}i,
\end{array}
\right.
\end{equation}
corresponding to a ``small'' \textit{critical} black string in $d=5$. This
configuration is the \textit{``electric-magnetic dual''} of the ``electric''
configuration \textbf{8}. However, differently from what happens for
configuration \textbf{8}, $\mathcal{PQ}\left( 2n_{V}+2,\mathbb{R}\right) $
is active (see Eq. (\ref{PQ-active})) in this case, due to its asymmetric
action on electric and magnetic charges. It generates $\Theta _{ij}$%
-dependent electric charges $q_{i}$'s and $\mathbf{c}_{i}$-dependent
electric charge $q_{0}$. The corresponding transformation of $\mathcal{I}%
_{4} $ reads
\begin{equation}
0\overset{\mathcal{O}^{-1}}{\longrightarrow }-\left( \Theta
_{ij}p^{i}p^{j}\right) ^{2}\leqslant 0.  \label{rell-6}
\end{equation}
Thus, according to (\ref{rell-5}), a ``small'' ($\mathcal{I}_{4}=0$:BPS or
non-BPS), or a ``large'' non-BPS $Z_{H}\neq 0$ ($\mathcal{I}_{4}<0$) BH
charge configuration can be generated. In case the quantity in (\ref{rell-6}%
) is strictly negative, this is an example of phenomenon (\ref{2.1-phen}).

\item  \textbf{``Small'' \textit{doubly-critical} (}$1$\textbf{-charge) }$%
p^{0}$ \textbf{(``magnetic'' Kaluza-Klein)} \textbf{configuration}. This is
the limit $q_{0}=0$ of configuration \textbf{1}. In Type $II$, it
corresponds to only $D6$ branes:
\begin{equation}
\mathcal{Q}\equiv \left( p^{0},0,0,0\right) ^{T}\Rightarrow \mathcal{I}%
_{4}\left( \mathcal{Q}\right) =0,
\end{equation}
The action of $\mathcal{PQ}\left( 2n_{V}+2,\mathbb{R}\right) $ reads
\begin{equation}
\left(
\begin{array}{l}
p^{0} \\
0 \\
0 \\
0
\end{array}
\right) \overset{\mathcal{O}^{-1}}{\longrightarrow }\left(
\begin{array}{l}
p^{0} \\
0 \\
-\varrho p^{0} \\
-\mathbf{c}_{i}p^{0}
\end{array}
\right) ,
\end{equation}
and thus it generates $\varrho $-dependent electric charge $q_{0}$ and $%
\mathbf{c}_{i}$-dependent electric charges $q_{i}$'s. These latter in Type $%
II$ compactifications corresponds to a stack of $D2$ branes depending on the
components of the second Chern class $c_{2}$ of $CY_{3}$ (recall Eq. (\ref
{2nd-Chern-class})). The corresponding transformation of $\mathcal{I}_{4}$
reads
\begin{equation}
0\overset{\mathcal{O}^{-1}}{\longrightarrow }\left( p^{0}\right) ^{4}\left(
\frac{2}{3}d^{ijk}\mathbf{c}_{i}\mathbf{c}_{j}\mathbf{c}_{k}-\varrho
^{2}\right) \gtreqless 0.
\end{equation}
Thus, depending on whether
\begin{equation}
\frac{2}{3}d^{ijk}\mathbf{c}_{i}\mathbf{c}_{j}\mathbf{c}_{k}-\varrho
^{2}\gtreqless 0,  \label{rell-7}
\end{equation}
a ``large'' ($\mathcal{I}_{4}>0$:BPS or non-BPS $Z_{H}=0$), a ``small'' ($%
\mathcal{I}_{4}=0$:BPS or non-BPS), or a ``large'' non-BPS $Z_{H}\neq 0$ ($%
\mathcal{I}_{4}<0$) BH charge configuration is generated. In case the
quantity in (\ref{rell-6}) is non-vanishing, this is an example of
phenomenon (\ref{2.1-phen}). Note that for $\varrho =0$, an example treated
in \cite{N=2-Quantum} is recovered, namely:
\begin{equation}
\left\{
\begin{array}{l}
0\overset{\mathcal{O}^{-1}}{\longrightarrow }4\left( p^{0}\right) ^{4}%
\mathcal{I}_{3}\left( \mathbf{c}\right) \gtreqless 0; \\
\\
\mathcal{I}_{3}\left( \mathbf{c}\right) \equiv \frac{1}{3!}d^{ijk}\mathbf{c}%
_{i}\mathbf{c}_{j}\mathbf{c}_{k}.
\end{array}
\right.
\end{equation}

\item  \textbf{``Small'' \textit{doubly-critical} (}$1$\textbf{-charge) }$%
q_{0}$ \textbf{(``electric'' Kaluza-Klein)} \textbf{configuration}. This is
the limit $p^{0}=0$ of configuration \textbf{1}. In Type $II$, it
corresponds to only $D0$ branes:
\begin{equation}
\mathcal{Q}\equiv \left( 0,0,q_{0},0\right) ^{T}\Rightarrow \mathcal{I}%
_{4}\left( \mathcal{Q}\right) =0,
\end{equation}
This configuration is the \textit{``electric-magnetic dual''} of the
``magnetic'' configuration \textbf{10}. Since there are no magnetic charges,
$\mathcal{PQ}\left( 2n_{V}+2,\mathbb{R}\right) $ is inactive on this
configuration:
\begin{equation}
\left(
\begin{array}{l}
0 \\
0 \\
q_{0} \\
0
\end{array}
\right) \overset{\mathcal{O}^{-1}}{\longrightarrow }\left(
\begin{array}{l}
0 \\
0 \\
q_{0} \\
0
\end{array}
\right) .
\end{equation}
\medskip
\end{enumerate}

We conclude this Sec. with a comment on the attractor values of the scalars,
\textit{i.e.} on the non-degenerate critical points of the effective BH
potential $V_{BH}$. In presence of the sub-leading quantum perturbative
corrections (\ref{F-sympl}), the expressions of such critical points can be
obtained from the ones for the uncorrected (not necessarily cubic) SK
geometry, by applying a suitable transformation of $\mathcal{PQ}\left(
2n_{V}+2,\mathbb{R}\right) $ on the charges.

This fact has been known for some time \cite{N=2-Quantum, Shmakova}. In the
case in which the uncorrected geometry is a $d$-SK geometry with
prepotential (\ref{d-SKG}), this provides a generally more efficient
approach to the computation of the attractor horizon (purely
charge-dependent) values of the scalars. Namely, one has to start from the
general expression of the extremal BH attractors for $d$-SK geometries \cite
{Shmakova,TT1}, and then apply the suitable transformation $\mathcal{O}^{-1}
$ (\ref{PQ-charge}) of $\mathcal{PQ}\left( 2n_{V}+2,\mathbb{R}\right) $ on
the charges. As an example, in this way the results recently obtained in
Sec. 3 and App. A of \cite{DT-1} can be recovered.

\subsubsection{\label{VBH-Transf}Transformation of $V_{BH}$}

As mentioned above, $\mathcal{PQ}\left( 2n_{V}+2,\mathbb{R}\right) $,
\textit{when acting both on the charges }$\mathcal{Q}$\textit{\ and on the
covariantly holomorphic symplectic sections }$\mathcal{V}$, leaves $Z$ and $%
D_{i}Z$, and thus $V_{BH}$ given by (\ref{VBH-def}), invariant.

Actually, in order to investigate the effect of the quantum perturbative
sub-leading corrections (\ref{F-sympl}) to any $\mathcal{N}=2$ prepotential
on $Z$, $D_{i}Z$, $V_{BH}$, $\partial _{i}V_{BH}$, $D_{i}\partial _{j}V_{BH}$%
, $D_{i}\overline{\partial }_{\overline{j}}V_{BH}$ etc., one should act with
$\mathcal{PQ}\left( 2n_{V}+2,\mathbb{R}\right) $ \textit{only on charges}.
In order to show this, let us consider (without any loss of generality for
our purposes) the $\mathcal{N}=2$ central charge $Z\equiv \left\langle
\mathcal{Q},\mathcal{V}\right\rangle \equiv \mathcal{Q}^{T}\Omega \mathcal{V}
$. By recalling that $\frak{F}$ can be introduced through the action of $%
\mathcal{O}\in \mathcal{PQ}\left( 2n_{V}+2,\mathbb{R}\right) $ (\ref{O}) on
the sections, the expression of $Z$ \textit{for any} $\mathcal{N}=2$
prepotential corrected with $\frak{F}$ (\ref{F-sympl}) is given by
\begin{eqnarray}
Z^{\prime } &\equiv &Z\left( \mathcal{OV}\left( z,\overline{z}\right) ;%
\mathcal{Q}\right) \equiv \left\langle \mathcal{Q},\mathcal{OV}\right\rangle
\equiv \mathcal{Q}^{T}\Omega \mathcal{OV}  \notag \\
&=&\mathcal{Q}^{T}\left( \mathcal{O}^{T}\right) ^{-1}\Omega \mathcal{V}%
=\left\langle \mathcal{O}^{-1}\mathcal{Q},\mathcal{V}\right\rangle \equiv
Z\left( \mathcal{V}\left( z,\overline{z}\right) ;\mathcal{O}^{-1}\mathcal{Q}%
\right) ,
\end{eqnarray}
where in the second line the symplectic nature of $\mathcal{O}$ has been
exploited. Thus, the expression of $Z$ \textit{for any} $\mathcal{N}=2$
prepotential corrected with $\frak{F}$ (\ref{F-sympl}) is nothing but the
expression of $Z$ computed for the uncorrected prepotential, with the
charges transformed through $\mathcal{O}$ given by (\ref{O}). The very same
holds also for $W,D_{i}W$, $D_{i}Z$, $V_{BH}$, $\partial _{i}V_{BH}$, $%
D_{i}\partial _{j}V_{BH}$, $D_{i}\overline{\partial }_{\overline{j}}V_{BH}$,
and in general for all quantities depending on scalars and charges. In the
case of the \textit{locus} $\partial _{i}V_{BH}=0$, this allows to easily
compute the $\frak{F}$-corrected attractors, once the ones for the
uncorrected prepotential are known (see the discussion at the end of Sec.
\ref{Analysis-Configs}). In the case in which the uncorrected SK geometry is
a cubic one, with prepotential (\ref{d-SKG}), this reasoning provides a
general alternative approach for the generalization (for all charge
configurations in which the treatment of the purely cubic case is feasible
\cite{Shmakova, TT1,CFM1}) of the computations recently performed in Sec. 3
and App. A of \cite{DT-1}.

In light of the previous reasoning, the explicit expressions of $Z$, $D_{i}Z$
and $V_{BH}$ for an $\frak{F}$-corrected $d$-SK geometry can be immediately
obtained by applying the charge transformation $\mathcal{O}^{-1}$ (given by (%
\ref{PQ-charge})) to Eqs. (4.9), (4.10) and (2.13) of \cite{CFM1},
respectively.

Since it is crucial to our treatment, we here consider only the $\frak{F}$%
-corrected expression of $V_{BH}$ for $d$-SK geometries. As mentioned, the
expression of $V_{BH}$ for $d$-SK geometries (\ref{d-SKG}) is given by Eq.
(2.13) of \cite{CFM1}, which we report here for ease of comparison:

\begin{eqnarray}
2V_{BH}\left( z,\overline{z};\mathcal{Q}\right) &=&\left[ \nu \left(
1+4g\right) +\frac{h^{2}}{36\nu }+\frac{3}{48\nu }g^{ij}h_{i}h_{j}\right]
\left( p^{0}\right) ^{2}+  \notag \\
&&+\left[ 4\nu g_{ij}+\frac{1}{4\nu }\left(
h_{i}h_{j}+g^{mn}h_{im}h_{nj}\right) \right] p^{i}p^{j}+  \notag \\
&&+\frac{1}{\nu }\left[ q_{0}^{2}+2x^{i}q_{0}q_{i}+\left( x^{i}x^{j}+\frac{1%
}{4}g^{ij}\right) q_{i}q_{j}\right] +  \notag \\
&&+2\left[ \nu g_{i}-\frac{h}{12\nu }h_{i}-\frac{1}{8\nu }g^{jm}h_{m}h_{ij}%
\right] p^{0}p^{i}+  \notag \\
&&-\frac{1}{3\nu }\left[
\begin{array}{l}
-hp^{0}q_{0}+3q_{0}p^{i}h_{i}-\left( hx^{i}+\frac{3}{4}g^{ij}h_{j}\right)
p^{0}q_{i} \\
\\
+3\left( h_{j}x^{i}+\frac{1}{2}g^{im}h_{mj}\right) q_{i}p^{j}
\end{array}
\right] ,  \label{V_BH-d-SKG}
\end{eqnarray}
where the following notation has been introduced (see e.g. \cite{CFM1} for
further details):
\begin{equation}
\left\{
\begin{array}{l}
z^{i}\equiv x^{i}-i\lambda ^{i}; \\
\\
\nu \equiv \frac{1}{3!}d_{ijk}\lambda ^{i}\lambda ^{j}\lambda ^{k}; \\
\\
h_{ij}\equiv d_{ijk}x^{k};~h_{i}\equiv d_{ijk}x^{j}x^{k};~h\equiv
d_{ijk}x^{i}x^{j}x^{k}; \\
\\
d_{ij}\equiv d_{ijk}\lambda ^{k};~d_{i}\equiv d_{ijk}\lambda ^{j}\lambda
^{k};~d^{ij}d_{jk}\equiv \delta _{k}^{i}; \\
\\
g_{ij}=-\frac{1}{4}\left( \frac{d_{ij}}{\nu }-\frac{d_{i}d_{j}}{4\nu ^{2}}%
\right) ;~g^{ij}=2\left( \lambda ^{i}\lambda ^{j}-2\nu d^{ij}\right) ; \\
\\
g_{i}\equiv -4g_{ij}x^{j};~g\equiv g_{ij}x^{i}x^{j}.
\end{array}
\right.  \label{d-SKG-notation}
\end{equation}
It is worth recalling that (\ref{V_BH-d-SKG}) was recently re-obtained as
the Im$d_{000}=0$ limit of the more general quantum perturbative result of
\cite{Raju-1}. Consistently with the above reasoning, straightforward
computations lead to the following expression of the $\frak{F}$-corrected
expression of $V_{BH}$ for $d$-SK geometries:
\begin{equation}
V_{BH}\left( z,\overline{z};\mathcal{Q}\right) \overset{\mathcal{O}^{-1}}{%
\longrightarrow }V_{BH}\left( z,\overline{z};\mathcal{O}^{-1}\mathcal{Q}%
\right) =V_{BH}\left( z,\overline{z};\mathcal{Q}\right) +\frak{V}_{BH}\left(
z,\overline{z};\mathcal{Q},\varrho ,\mathbf{c}_{i},\Theta _{ij}\right) ,
\label{V-transf}
\end{equation}
where $\frak{V}_{BH}$ describes the \textit{``PQ-deformation''} of $V_{BH}$:
\begin{eqnarray}
2\frak{V}_{BH}\left( z,\overline{z};\mathcal{Q},\varrho ,\mathbf{c}%
_{i},\Theta _{ij}\right) &=&\frac{1}{\nu }\left[
\begin{array}{l}
\varrho ^{2}\left( p^{0}\right) ^{2}+\left( \mathbf{c}_{i}p^{i}\right)
^{2}-2q_{0}\varrho p^{0}-2q_{0}\mathbf{c}_{i}p^{i}+2\varrho p^{0}\mathbf{c}%
_{i}p^{i} \\
\\
+2x^{i}\left(
\begin{array}{l}
-p^{0}q_{0}\mathbf{c}_{i}-q_{0}\Theta _{ij}p^{j} \\
\\
-\varrho p^{0}q_{i}+\varrho \left( p^{0}\right) ^{2}\mathbf{c}_{i}+\varrho
p^{0}\Theta _{ij}p^{j} \\
\\
-\mathbf{c}_{j}p^{j}q_{i}+p^{0}\mathbf{c}_{i}\mathbf{c}_{j}p^{j}+\mathbf{c}%
_{j}p^{j}\Theta _{ik}p^{k}
\end{array}
\right) \\
\\
+\left( x^{i}x^{j}+\frac{1}{4}g^{ij}\right) \left(
\begin{array}{l}
-q_{i}\mathbf{c}_{j}p^{0}-q_{i}\Theta _{jl}p^{l} \\
\\
-p^{0}\mathbf{c}_{i}q_{j}+\left( p^{0}\right) ^{2}\mathbf{c}_{i}\mathbf{c}%
_{j}+p^{0}\mathbf{c}_{i}\Theta _{jk}p^{k} \\
\\
-\Theta _{ik}p^{k}q_{j}+p^{0}\Theta _{ik}p^{k}\mathbf{c}_{j}+\Theta
_{ik}p^{k}\Theta _{jl}p^{l}
\end{array}
\right)
\end{array}
\right] +  \notag \\
&&  \notag \\
&&-\frac{1}{3\nu }\left[
\begin{array}{l}
hp^{0}\left( \varrho p^{0}+\mathbf{c}_{i}p^{i}\right) \\
\\
-3\left( \varrho p^{0}+\mathbf{c}_{j}p^{j}\right) p^{i}h_{i} \\
\\
+\left( hx^{i}+\frac{3}{4}g^{ij}h_{j}\right) p^{0}\left( \mathbf{c}%
_{i}p^{0}+\Theta _{ik}p^{k}\right) \\
\\
-3\left( h_{j}x^{i}+\frac{1}{2}g^{im}h_{mj}\right) \left( \mathbf{c}%
_{i}p^{0}+\Theta _{ik}p^{k}\right) p^{j}
\end{array}
\right] .  \notag \\
&&  \label{V-Frak}
\end{eqnarray}

Eqs. (\ref{V-transf}), (\ref{V_BH-d-SKG}) and (\ref{V-Frak}), once specified
for the particular $n_{V}=2$ model treated in \cite{DT-1} (see Eq. (3.7)
therein), allows one to easily recover Eq. (A.12) therein. Furthermore, by
setting $p^{0}=0=q_{i}$ (\textit{i.e.} by considering the $D0-D4$
configuration), Eq. (\ref{V-Frak}) yields that the $\frak{F}$-corrected $%
V_{BH}$ does not depend at all on $\varrho $; this fact generalizes the
comment below Eq. (3.1) of \cite{DT-1}.\medskip

Let us now consider the part of $V_{BH}$ (\ref{V_BH-d-SKG}) linear in the
axions $x^{i}$. Eq. (\ref{V_BH-d-SKG}) yields
\begin{equation}
\left. 2V_{BH}\right| _{\text{linear~in~}x^{i}}=\frac{2}{\nu }%
x^{i}q_{0}q_{i}+2\nu g_{i}p^{0}p^{i}-\frac{1}{2\nu }g^{ik}h_{kj}q_{i}p^{j}.
\label{pre-PQ-linear-x}
\end{equation}
This implies that the BH charge configurations which support the \textit{%
axion-free} solution $x^{i}=0~\forall i$ \textit{at least} as a particular
solution of the axionic Attractor Eqs. $\partial V_{BH}/\partial x^{i}=0$
are the following ones \cite{CFM1}:
\begin{equation}
\left\{
\begin{array}{l}
\left( p^{0},q_{0}\right) ; \\
\left( p^{0},q_{i}\right) ; \\
\left( p^{i},q_{0}\right) ,
\end{array}
\right.  \label{pre-PQ-axion-free}
\end{equation}
namely the ``large'' configurations \textbf{1}, \textbf{2} and \textbf{3}
treated in Sec. \ref{Analysis-Configs}.

Through Eqs. (\ref{V-transf}), (\ref{V_BH-d-SKG}) and (\ref{V-Frak}), the
action of $\mathcal{PQ}\left( 2n_{V}+2,\mathbb{R}\right) $ transforms (\ref
{pre-PQ-linear-x}) as follows:
\begin{eqnarray}
\left. 2\left[ V_{BH}+\frak{V}_{BH}\right] \right| _{\text{linear~in~}x^{i}}
&=&\frac{2}{\nu }x^{i}q_{0}q_{i}+2\nu g_{i}p^{0}p^{i}-\frac{1}{2\nu }%
g^{ik}h_{kj}q_{i}p^{j}  \notag \\
&&+\frac{2}{\nu }x^{i}\left(
\begin{array}{l}
-p^{0}q_{0}\mathbf{c}_{i}-q_{0}\Theta _{ij}p^{j} \\
\\
-\varrho p^{0}q_{i}+\varrho \left( p^{0}\right) ^{2}\mathbf{c}_{i}+\varrho
p^{0}\Theta _{ij}p^{j} \\
\\
-\mathbf{c}_{j}p^{j}q_{i}+p^{0}\mathbf{c}_{i}\mathbf{c}_{j}p^{j}+\mathbf{c}%
_{j}p^{j}\Theta _{ik}p^{k}
\end{array}
\right)  \notag \\
&&+\frac{1}{2\nu }g^{im}h_{mj}\left( \mathbf{c}_{i}p^{0}+\Theta
_{ik}p^{k}\right) p^{j}.  \label{post-PQ-linear-x}
\end{eqnarray}
The rather intricate expression (\ref{post-PQ-linear-x}) implies that, in
presence of the sub-leading quantum perturbative corrections (\ref{F-sympl}%
), the configurations (\ref{pre-PQ-axion-free}) do not support axion-free
solutions any more, and that in general there are no axion-free-supporting
BH charge configurations at all\footnote{%
This result is consistent with the analysis of the particular $n_{V}=2$
model in $D0$-$D4$ configuration worked out in \cite{DT-1}.}, \textit{unless
some extra assumptions are made}. For instance, (\ref{post-PQ-linear-x})
yields the following axion-free-supporting conditions for the charge
configurations (\ref{pre-PQ-axion-free}):
\begin{equation}
\left. 2\left[ V_{BH}+\frak{V}_{BH}\right] \right| _{\text{linear~in~}%
x^{i},\left( p^{0},q_{0}\right) }=\frac{2}{\nu }x^{i}\mathbf{c}%
_{i}p^{0}\left( -q_{0}+\varrho p^{0}\right) =0\Leftrightarrow \left\{
\begin{array}{l}
\mathbf{c}_{i}=0; \\
\text{\textit{and/or}} \\
q_{0}=\varrho p^{0};
\end{array}
\right.  \label{D0-D6-1}
\end{equation}
\begin{equation}
\left. 2\left[ V_{BH}+\frak{V}_{BH}\right] \right| _{\text{linear~in~}%
x^{i},\left( p^{0},q_{i}\right) }=\frac{2}{\nu }x^{i}\varrho p^{0}\left(
-q_{i}+p^{0}\mathbf{c}_{i}\right) =0\Leftrightarrow \left\{
\begin{array}{l}
\varrho =0; \\
\text{\textit{and/or}} \\
q_{i}=p^{0}\mathbf{c}_{i};
\end{array}
\right.
\end{equation}
\begin{gather}
\left. 2\left[ V_{BH}+\frak{V}_{BH}\right] \right| _{\text{linear~in~}%
x^{i},\left( p^{i},q_{0}\right) }=\frac{2}{\nu }\left( -\delta
_{m}^{i}q_{0}+\delta _{m}^{i}\mathbf{c}_{k}p^{k}+\frac{1}{4}%
g^{il}d_{klm}p^{k}\right) \Theta _{ij}p^{j}x^{m}=0  \notag \\
\Updownarrow  \notag \\
\left\{
\begin{array}{l}
\Theta _{ij}=0; \\
\text{\textit{and/or}} \\
-\delta _{m}^{i}q_{0}+\delta _{m}^{i}\mathbf{c}_{k}p^{k}+\frac{1}{4}%
g^{il}d_{klm}p^{k}=0.
\end{array}
\right.
\end{gather}

It is known \cite{Ferrara-Marrani-2} that in symmetric $d$-SK geometries,
the \textit{``moduli space''} of non-BPS $Z_{H}\neq 0$ attractors is the
scalar manifold of the $d=5$ uplifted theory. This can be easily seen in the
$\left( p^{0},q_{0}\right) $ configuration. Indeed, by setting $x^{i}=0$ $%
\forall i$, the effective BH potential (\ref{V_BH-d-SKG}) reads
\begin{equation}
2\left. V_{BH}\right| _{\left( p^{0},q_{0}\right) ,x^{i}=0~\forall i}=\nu
\left( p^{0}\right) ^{2}+\frac{1}{\nu }q_{0}^{2},  \label{pre-PQ-crit}
\end{equation}
thus depending only on the Kaluza-Klein volume $\nu $. The $n_{V}$ real
``rescaled'' dilatons \cite{CFM1}
\begin{equation}
\widehat{\lambda }^{i}\equiv \nu ^{-\frac{1}{3}}\lambda ^{i},
\label{lambda-hat-def}
\end{equation}
which defines the $d=5$ scalar manifold through the cubic constraint
\begin{equation}
\frac{1}{3!}d_{ijk}\widehat{\lambda }^{i}\widehat{\lambda }^{i}\widehat{%
\lambda }^{i}=1  \label{lambda-hat-propr}
\end{equation}
are \textit{``flat directions''} of the critical value (\ref{pre-PQ-crit}).

The action of $\mathcal{PQ}\left( 2n_{V}+2,\mathbb{R}\right) $ may make the
emergence of \textit{``moduli spaces''} of attractors less manifest but, as
stated above, does not change their geometrical structure. From (\ref
{D0-D6-1}), in $\frak{F}$-corrected $d$-SK geometry the Kaluza-Klein charge
configuration $\left( p^{0},q_{0}\right) $ (with no further constraints) is
axion-free-supporting for $\mathbf{c}_{i}=0$ $\forall i$. In such a case,
Eqs. (\ref{D0-D6-1}) and (\ref{pre-rell}) respectively yield
\begin{equation}
\left. 2\left[ V_{BH}+\frak{V}_{BH}\right] \right| _{\left(
p^{0},q_{0}\right) ,x^{i}=0~\forall i}=\nu \left( p^{0}\right) ^{2}+\frac{1}{%
\nu }\left( q_{0}-\varrho p^{0}\right) ^{2};  \label{post-PQ-crit}
\end{equation}
\begin{equation}
-\left( p^{0}\right) ^{2}q_{0}^{2}\overset{\mathcal{O}^{-1}}{\longrightarrow
}-\left( p^{0}\right) ^{2}\left( q_{0}-\varrho p^{0}\right) ^{2}.
\end{equation}
Thus, the PQ-transformed BH charge configuration $\left( p^{0},q_{0}\right) $
with $\mathbf{c}_{i}=0$ $\forall i$ (and $q_{0}\neq \varrho p^{0}$) still
supports non-BPS $Z_{H}\neq 0$ (possibly axion-free) extremal BH attractors,
whose \textit{``moduli space''} is still manifest from (\ref{post-PQ-crit}).
Note that the case $q_{0}=\varrho p^{0}$ is troublesome, because it does not
stabilize the Kaluza-Klein volume through the Attractor Mechanism.

\subsection{\label{PQ-Ell-Cayley}Cayley's Hyperdeterminant and Elliptic
Curves}

Recently, in \cite{G-1}, an intriguing relation between elliptic curves and
the Cayley's hyperdeterminant \cite{Cayley} was found.

More specifically, it was shown that if the cubic elliptic curve
\begin{equation}
y^{2}=ax^{3}+bx^{2}+cx+d  \label{1-ell}
\end{equation}
has a Mordell-Weil group containing a subgroup isomorphic to $\mathbb{Z}%
\times \mathbb{Z}_{2}\times \mathbb{Z}_{2}$, then it can be transformed into
the Cayley's hyperdeterminant Det$\left( \psi \right) $, which is nothing
but the (opposite of the) quartic scalar invariant built out of the unique
rank-$4$ completely symmetric primitive invariant tensor of the repr. $%
\left( \mathbf{2},\mathbf{2},\mathbf{2}\right) $ of $\left[ SL\left( 2,%
\mathbb{R}\right) \right] ^{3}$, which in turn is the $U$-duality group of
the $\mathcal{N}=2$, $d=4$ so-called $stu$ model \cite{stu}:
\begin{eqnarray}
\mathcal{I}_{4,stu}\left( \mathcal{Q}\right) &=&-\left( p^{0}\right)
^{2}q_{0}^{2}-\left( p^{1}\right) ^{2}q_{1}^{2}-\left( p^{2}\right)
^{2}q_{2}^{2}-\left( p^{3}\right) ^{2}q_{3}^{2}  \notag \\
&&-2p^{0}q_{0}p^{1}q_{1}-2p^{0}q_{0}p^{2}q_{2}-2p^{0}q_{0}p^{3}q_{3}+2p^{1}q_{1}p^{2}q_{2}+2p^{1}q_{1}p^{3}q_{3}+2p^{2}q_{2}p^{3}q_{3}
\notag \\
&&+4q_{0}p^{1}p^{2}p^{3}-4p^{0}q_{1}q_{2}q_{3}=-\text{Det}\left( \psi
\right) .  \label{I4-stu}
\end{eqnarray}
This expression can be obtained from the general one (\ref{I4-bare-1})-(\ref
{defs-I4-bare}), by specifying the $stu$ model data:
\begin{equation}
d_{ijk}=6\delta _{1\left( i\right| }\delta _{2\left| j\right| }\delta
_{3\left| k\right) };~d^{ijk}=6\delta ^{1\left( i\right| }\delta ^{2\left|
j\right| }\delta ^{3\left| k\right) },  \label{stu-data}
\end{equation}
consistent with (\ref{id-symm}).

Under the aforementioned assumption on the Mordell-Weil group, the elliptic
curve (\ref{1-ell}) can be factorised as \cite{G-1}
\begin{equation}
y^{2}=4\left( l-kx\right) \left( n-mx\right) \left( q-px\right) ,
\label{factor}
\end{equation}
and through the positions (with $u$, $v$ unknowns) \cite{G-1}
\begin{eqnarray}
y &=&uv^{2}-ev+g;  \label{pos-1} \\
x &=&v;  \label{pos-2} \\
a &=&-4kmp;  \label{pos-3} \\
b &=&4kmrt+4kpts+4mprs;  \label{pos-4} \\
c &=&-4rts\left( kt+mr+ps\right) ;  \label{pos-5} \\
d &=&4r^{2}s^{2}t^{2},  \label{pos-6}
\end{eqnarray}
finally (\ref{1-ell}) can be recast in the form
\begin{equation}
u^{2}v^{2}+k^{2}t^{2}+m^{2}r^{2}+p^{2}s^{2}-2ktuv-2mruv-2psuv-2kmrt-2kpts-2mprs+4kmpv+4rstu=0,
\label{2-2}
\end{equation}
which corresponds to the vanishing of $\mathcal{I}_{4,stu}\left( \mathcal{Q}%
\right) $ as given by (\ref{I4-stu}), under the (non-unique) following
mapping of the charge vector:
\begin{equation}
\mathcal{Q}\equiv \left(
p^{0},p^{1},p^{2},p^{3},q_{0},q_{1},q_{2},q_{3}\right) ^{T}=\left(
u,k,m,p,-v,t,r,s\right) ^{T}.  \label{non-unique-mapping}
\end{equation}
Interestingly, the two unknowns $u$ and $v$ corresponds to the magnetic ($D6$%
) and electric ($D0$) Kaluza-Klein charges in the reduction $d=5\rightarrow
d=4$.

Under the position (\ref{non-unique-mapping}), the vanishing of $\mathcal{I}%
_{4,stu}\left( \mathcal{Q}\right) $, a necessary condition defining the
``small'' orbits of the $\left( \mathbf{2},\mathbf{2},\mathbf{2}\right) $ of
$\left[ SL\left( 2,\mathbb{R}\right) \right] ^{3}$ \cite{Duff-stu-small},
can be recast in the form (\ref{1-ell}), with
\begin{eqnarray}
y &=&p^{0}q_{0}^{2}+q_{0}\left( p^{1}q_{1}+p^{2}q_{2}+p^{3}q_{3}\right)
+2q_{1}q_{2}q_{3};  \label{charges-y} \\
x &=&-q_{0}; \\
a &=&-4p^{1}p^{2}p^{3}; \\
b &=&4\left(
p^{1}q_{1}p^{2}q_{2}+p^{1}q_{1}p^{3}q_{3}+p^{2}q_{2}p^{3}q_{3}\right) ; \\
c &=&-4q_{1}q_{2}q_{3}\left( p^{1}q_{1}+p^{2}q_{2}+p^{3}q_{3}\right) ; \\
d &=&4q_{1}^{2}q_{2}^{2}q_{3}^{2},  \label{charges-d}
\end{eqnarray}

In light of the treatment given in Secs. \ref{GT} and \ref{Applications}, it
is worth pointing out that the above construction admits a ``$\mathcal{PQ}%
\left( 8,\mathbb{R}\right) $-deformation''.

The ``$\mathcal{PQ}\left( 8,\mathbb{R}\right) $-deformation'' of the
Cayley's hyperdeterminant can be obtained from the general result (\ref
{PQ-I4})-(\ref{PQ-Ibold4}) by using the $stu$ model data (\ref{stu-data})
(here $i,j=1,2,3$):
\begin{eqnarray}
&&\mathcal{I}_{4,stu}\left( \mathcal{Q}\right) +\mathbf{I}_{4.stu}\left(
\mathcal{Q};\varrho ,\mathbf{c}_{i},\Theta _{ij}\right) =-\left(
p^{0}\right) ^{2}\left( q_{0}-\varrho p^{0}-\mathbf{c}_{i}p^{i}\right)
^{2}-\left( p^{i}\right) ^{2}\left( q_{i}-\mathbf{c}_{i}p^{0}-\Theta
_{ij}p^{j}\right) ^{2}  \notag \\
&&-2p^{0}p^{i}\left( q_{0}-\varrho p^{0}-\mathbf{c}_{j}p^{j}\right) \left(
q_{i}-\mathbf{c}_{i}p^{0}-\Theta _{ij}p^{j}\right)  \notag \\
&&+\sum_{i=1}^{3}\left| \epsilon _{ijk}\right| p^{j}\left( q_{j}-\mathbf{c}%
_{j}p^{0}-\Theta _{jl}p^{l}\right) p^{k}\left( q_{k}-\mathbf{c}%
_{k}p^{0}-\Theta _{km}p^{m}\right)  \notag \\
&&+4\left( q_{0}-\varrho p^{0}-\mathbf{c}_{i}p^{i}\right)
p^{1}p^{2}p^{3}-4p^{0}\prod_{i=1}^{3}\left( q_{i}-\mathbf{c}_{i}p^{0}-\Theta
_{ij}p^{i}\right)  \notag \\
&=&-\text{Det}\left( \psi ;\varrho ,\mathbf{c}_{i},\Theta _{ij}\right) .
\end{eqnarray}
The various terms (unknowns and coefficients) of the corresponding cubic
elliptic curve (\ref{1-ell}) are given by the $\mathcal{PQ}\left( 8,\mathbb{R%
}\right) $-transformed Eqs. (\ref{charges-y})-(\ref{charges-d}), namely:
\begin{eqnarray}
y &=&p^{0}\left( q_{0}-\varrho p^{0}-\mathbf{c}_{i}p^{i}\right) ^{2}+\left(
q_{0}-\varrho p^{0}-\mathbf{c}_{i}p^{i}\right) p^{i}\left( q_{i}-\mathbf{c}%
_{i}p^{0}-\Theta _{ij}p^{j}\right)  \notag \\
&&+2\prod_{i}\left( q_{i}-\mathbf{c}_{i}p^{0}-\Theta _{ij}p^{i}\right) ;
\label{charges-y-PQ} \\
x &=&-\left( q_{0}-\varrho p^{0}-\mathbf{c}_{i}p^{i}\right) ; \\
a &=&-4p^{1}p^{2}p^{3}; \\
b &=&2\sum_{i}\left| \epsilon _{ijk}\right| p^{j}\left( q_{j}-\mathbf{c}%
_{j}p^{0}-\Theta _{jl}p^{l}\right) p^{k}\left( q_{k}-\mathbf{c}%
_{k}p^{0}-\Theta _{km}p^{m}\right) ; \\
c &=&-2\sum_{k}\left| \epsilon _{klm}\right| p^{l}\left( q_{l}-\mathbf{c}%
_{l}p^{0}-\Theta _{ln}p^{n}\right) p^{m}\left( q_{m}-\mathbf{c}%
_{m}p^{0}-\Theta _{mr}p^{r}\right) \prod_{i}\left( q_{i}-\mathbf{c}%
_{i}p^{0}-\Theta _{ij}p^{i}\right) ;  \notag \\
&& \\
d &=&4\left[ \prod_{i}\left( q_{i}-\mathbf{c}_{i}p^{0}-\Theta
_{ij}p^{i}\right) \right] ^{2}.  \label{charges-d-PQ}
\end{eqnarray}

Clearly, the roots of the elliptic cubic curve (\ref{1-ell}) (with data (\ref
{charges-y})-(\ref{charges-d})) are not the same as the roots of (\ref{1-ell}%
) (with data (\ref{charges-y-PQ})-(\ref{charges-d-PQ})). In general, the
action of $\mathcal{PQ}\left( 8,\mathbb{R}\right) $ amounts to a $\left(
\varrho ,\mathbf{c}_{i},\Theta _{ij}\right) $-redefinition of the vertices
of the hypercube whose associate hyperdeterminant is Det$\left( \psi \right)
$ given by (\ref{I4-stu}) \cite{Cayley}.

Let us further remark that in realistic superstring compactifications
leading to the $stu$ model in the supergravity limit, the values of the
parameters $\mathbf{c}_{i}=\frac{c_{2,i}}{24}$ (where $c_{2}$ is the second
Chern class; see Sec. \ref{Stringy}) can be computed to read \cite
{0704.2440,0603149}:
\begin{equation}
\begin{array}{ll}
\text{Type~}IIA\text{~on~}K3~\text{fibrations} &
:c_{2,1}=c_{2,3}=24;~c_{2,2}=92; \\
\text{Heterotic~on~}T^{4}\times T^{2}~\text{or~}K3\times T^{2} &
:c_{2,2}=c_{2,3}=0.
\end{array}
\end{equation}

In view of the recent progress within the fascinating BH/qubit
correspondence \cite{QIT}, $\mathcal{PQ}\left( 8,\mathbb{R}\right) $ may
well have a role on the quantum information side; we leave the study of this
interesting issue for future investigation.

\section{\label{Alternative-I4}An Alternative Expression for $\mathcal{I}%
_{4} $}

By refining and extending the analysis of \cite{Raju-1} and considering $d$%
-SK geometries based on the purely cubic holomorphic prepotential (\ref
{d-SKG}), we will now derive an alternative expression of the quartic
invariant $\mathcal{I}_{4}$ given by (\ref{I4-bare-1})-(\ref{defs-I4-bare}).

A crucial quantity in such developments is the so-called $E$-tensor. Such a
rank-$5$ tensor was firstly introduced in \cite{dWVVP} (see also the
treatment of \cite{CVP}), and it expresses the deviation of the considered
geometry from being symmetric. Its definition reads (see \textit{e.g.} \cite
{Kallosh-rev,Raju-1} for a recent treatment, and Refs. therein):
\begin{equation}
\overline{E}_{\overline{m}ijkl}\equiv \frac{1}{3}\overline{D}_{\overline{m}%
}D_{i}C_{jkl}.  \label{E}
\end{equation}
This definition can be elaborated further, by recalling the properties of
the so-called $C$-tensor $C_{ijk}$. This is a rank-$3$ tensor with
K\"{a}hler weights $\left( 2,-2\right) $, defined as (see \textit{e.g.} \cite
{CDF-rev,Castellani1}):
\begin{eqnarray}
C_{ijk} &\equiv &\left\langle D_{i}D_{j}\mathbf{V},D_{k}\mathbf{V}%
\right\rangle =e^{K}\left( \partial _{i}\mathcal{N}_{\Lambda \Sigma }\right)
D_{j}X^{\Lambda }D_{k}X^{\Sigma }  \notag \\
&=&e^{K}\left( \partial _{i}X^{\Lambda }\right) \left( \partial
_{j}X^{\Sigma }\right) \left( \partial _{k}X^{\Xi }\right) \partial _{\Xi
}\partial _{\Sigma }F_{\Lambda }\left( X\right) \equiv e^{K}W_{ijk},\text{~~}%
\overline{\partial }_{\overline{l}}W_{ijk}=0,  \label{C}
\end{eqnarray}
where $\mathcal{N}_{\Lambda \Sigma }$ is the $\mathcal{N}=2$, $d=4$ kinetic
vector matric, and the second line holds only in ``special coordinates''. $%
C_{ijk}$ is completely symmetric and covariantly holomorphic:
\begin{equation}
C_{ijk}=C_{\left( ijk\right) };~~\overline{D}_{\overline{i}}C_{jkl}=0.
\end{equation}
By further steps, detailed in \cite{Raju-1}, the expression for $\overline{E}%
_{\overline{m}ijkl}$ defined by (\ref{E}) can thus be further elaborated as
follows:
\begin{equation}
C_{p(ij}C_{kl)q}g^{p\overline{r}}g^{q\overline{s}}\overline{C}_{\overline{r}%
\overline{s}\overline{t}}=\frac{4}{3}C_{(ijk}g_{l)\overline{t}}+\overline{E}%
_{\overline{t}ijkl}.  \label{E-elab}
\end{equation}

Formul\ae\ (\ref{E}) and (\ref{E-elab}) hold for a generic SK geometry. By
considering $d$-SK geometries based on the purely cubic holomorphic
prepotential (\ref{d-SKG}) in the ``special coordinates'' symplectic basis, (%
\ref{E-elab}) can be recast as
\begin{equation}
\left( X^{0}\right) ^{3}e^{3\mathcal{K}}d_{p(ij}d_{kl)q}g^{pr}g^{qs}d_{rst}=%
\frac{4}{3}X^{0}e^{\mathcal{K}}d_{(ijk}g_{l)t}+E_{tijkl},
\end{equation}
where $g^{ij}$ and $g_{ij}$ are defined in (\ref{d-SKG-notation}) (see
\textit{e.g.} \cite{CFM1} for further details).

Let us now introduce the ``rescaled metric'' \cite{FG-d=5,CFM1} and, for
later convenience, its derivatives with respect to $\widehat{\lambda }^{i}$
(the unique set of scalars on which it actually depends):
\begin{eqnarray}
a_{ij} &\equiv &4\nu ^{2/3}g_{ij}=\left( \frac{1}{4}\widehat{d}_{i}\widehat{d%
}_{j}-\widehat{d}_{ij}\right) \Leftrightarrow a^{ij}=\frac{1}{4}\nu
^{-2/3}g^{ij}=\frac{1}{2}\widehat{\lambda }^{i}\widehat{\lambda }^{j}-%
\widehat{d}^{ij};  \label{a^ij-def} \\
\frac{\partial a_{ij}}{\partial \widehat{\lambda }^{k}} &=&\frac{1}{2}\left(
\widehat{d}_{ik}\widehat{d}_{j}+\widehat{d}_{jk}\widehat{d}_{i}\right)
-d_{ijk};  \label{a_ij-diff} \\
\frac{\partial a^{ij}}{\partial \widehat{\lambda }^{k}} &=&\frac{1}{2}\left(
\delta _{k}^{i}\widehat{\lambda }^{j}+\delta _{k}^{j}\widehat{\lambda }%
^{i}\right) +\widehat{d}^{il}\widehat{d}^{jm}d_{klm},  \label{a^ij-diff}
\end{eqnarray}
where $\widehat{d}_{ij}$, $\widehat{d}^{ij}$ and $\widehat{d}_{i}$ are the
``hatted'' counterpart of the quantities defined in (\ref{d-SKG-notation})
(also recall the splitting $z^{i}\equiv x^{i}-i\lambda ^{i}$ in the first
line of (\ref{d-SKG-notation}), as well as (\ref{lambda-hat-def}) and (\ref
{lambda-hat-propr})):
\begin{eqnarray}
\widehat{d}_{ij} &\equiv &d_{ijk}\widehat{\lambda }^{k};~\widehat{d}%
_{i}\equiv d_{ijk}\widehat{\lambda }^{j}\widehat{\lambda }^{k}; \\
\widehat{d}^{ij}\widehat{d}_{jk} &\equiv &\delta _{k}^{i}\Rightarrow \frac{%
\partial \widehat{d}^{im}}{\partial \widehat{\lambda }^{k}}=-\widehat{d}^{ij}%
\widehat{d}^{ml}d_{jkl}.
\end{eqnarray}
Thus, by fixing the K\"{a}hler gauge $X^{0}\equiv 1$, after some algebra one
achieves the following result\footnote{%
Note that in $d$-SK geometries all geometrical quantities under
consideration are real.}:
\begin{equation}
d_{p(ij}d_{kl)q}d^{pqv}=\frac{4}{3}\delta _{(l}^{v}d_{ijk)}+2^{5}\nu
^{5/3}E_{~ijkl}^{v},  \label{result-1}
\end{equation}
where
\begin{eqnarray}
d^{pqv} &\equiv &a^{pr}a^{qs}a^{vt}d_{rst};  \label{d^ijk-def} \\
E_{~ijkl}^{v} &\equiv &a^{vt}E_{tijkl}.
\end{eqnarray}
From (\ref{result-1}), one can re-derive the explicit expression of $%
E_{tijkl}$ given by Eq. (4.21) of\footnote{%
For homogeneous non-symmetric $d$-SK geometries, the expression of the $E$%
-tensor was explicitly computed in \cite{ADT-Hom-Non-Symm}.} \cite{Raju-1},
implying that in any $d$-SK geometry $\nu ^{5/3}E_{tijkl}$ depends only on
the ``rescaled $d=4$ dilatons'' $\widehat{\lambda }^{i}$.

Let us now introduce the following $p^{i}$-dependent quantities, which are
scalar-independent in any $d$-SK geometry\footnote{%
Attention should be paid not to confuse the scalar-independent quantities $%
\mathbf{d}_{ij}$ and $\mathbf{d}^{ij}$ defined by (\ref{dbold(p)}) with the $%
\lambda ^{i}$-dependent quantities $d_{ij}$ and $d_{ij}$ defined in (\ref
{d-SKG-notation}).}:
\begin{equation}
\mathbf{d}_{ij}\equiv d_{ijk}p^{k}=\frac{\partial \mathcal{I}_{3}\left(
p\right) }{\partial p^{i}\partial p^{j}};~\mathbf{d}^{ij}\mathbf{d}%
_{jk}\equiv \delta _{k}^{i},  \label{dbold(p)}
\end{equation}
from which the following behaviors follow: $\mathbf{d}_{ij}\sim \left[ p%
\right] ^{2}$ and $\mathbf{d}^{ij}\sim \left[ p\right] ^{-2}$.

Thus, whenever $\mathbf{d}_{ij}$ has maximal rank $n_{V}$, by contracting (%
\ref{result-1}) with $p^{k}p^{l}p^{i}q_{v}q_{t}\mathbf{d}^{jt}$, a little
algebra leads to the result
\begin{equation}
-\left( p^{i}q_{i}\right) ^{2}+d_{ijk}d^{ilm}p^{j}p^{k}q_{l}q_{m}=\frac{1}{3}%
d_{ijk}p^{i}p^{j}p^{k}q_{l}q_{m}\mathbf{d}^{lm}+2^{5}\nu
^{5/3}E_{~ijkl}^{m}p^{j}p^{k}p^{l}q_{m}q_{n}\mathbf{d}^{in}.
\label{result-2}
\end{equation}
By plugging (\ref{result-2}) into the general expression of $\mathcal{I}_{4}$
given by (\ref{I4-bare-1})-(\ref{defs-I4-bare}), one obtains the following
alternative expression:
\begin{eqnarray}
\mathcal{I}_{4} &=&-\left( p^{0}\right) ^{2}q_{0}^{2}-2p^{0}q_{0}p^{i}q_{i}+%
\frac{1}{3}\left( 2q_{0}+q_{i}q_{j}\mathbf{d}^{ij}\right)
d_{klm}p^{k}p^{l}p^{m}  \notag \\
&&-\frac{2}{3}p^{0}d^{ijk}q_{i}q_{j}q_{k}+2^{5}\nu
^{5/3}E_{~ijkl}^{m}p^{j}p^{k}p^{l}q_{m}q_{n}\mathbf{d}^{in},\label{res-I4}
\end{eqnarray}
which manifestly shows the contribution of the $E$-tensor as a source of
dependence on $\nu $ and $\widehat{\lambda }^{i}$'s for non-symmetric $d$-SK
geometries, and more in general for all $d$-SK geometries in which the term $%
E_{~ijkl}^{m}p^{j}p^{k}p^{l}q_{m}q_{n}\mathbf{d}^{in}$ does not
vanish. Note that (\ref{res-I4}) is well defined whenever
$\mathbf{d}_{ij}$ (introduced in (\ref{dbold(p)})) has maximal rank
$n_{V}$.

Some comments on the alternative formula (\ref{res-I4}) for $\mathcal{I}_{4}$
are in order.

\begin{enumerate}
\item  In \textit{symmetric} $d$-SK geometries (see \textit{e.g.} \cite
{CVP,dWVVP}, and Refs. therein) $E_{mijkl}=0$, as a consequence of the
covariant constancy of the Riemann tensor $R_{i\overline{j}k\overline{l}}$
itself (see \textit{e.g.} \cite{Raju-1} for a recent treatment):
\begin{equation}
D_{m}R_{i\overline{j}k\overline{l}}=0.  \label{DR=0}
\end{equation}
This implies, through Eq. (\ref{E-elab}):
\begin{equation}
C_{p(kl}C_{ij)n}g^{n\overline{n}}g^{p\overline{p}}\overline{C}_{\overline{n}%
\overline{p}\overline{m}}=\frac{4}{3}g_{\left( l\right| \overline{m}%
}C_{\left| ijk\right) }\Leftrightarrow g^{n\overline{n}}R_{\left( i\right|
\overline{m}\left| j\right| \overline{n}}C_{n\left| kl\right) }=-\frac{2}{3}%
g_{\left( i\right| \overline{m}}C_{\left| jkl\right) },  \label{symm}
\end{equation}
whose specification in the manifestly $G_{5}$-covariant ``special
coordinates'' symplectic basis gives the identity (\ref{id-symm}), which is
consistently the $E_{~ijkl}^{v}=0$ limit of (\ref{result-1}). By recalling
definition (\ref{d^ijk-def}), (\ref{id-symm}) (holding for symmetric $d$%
-SKG, and more in general in all cases in which $E_{~ijkl}^{v}=0$ globally)
implies that $d^{ijk}$ is a constant, scalar-independent tensor:
\begin{equation}
\frac{\partial d^{ijk}}{\partial z^{l}}=0.
\end{equation}
Furthermore, the $E_{~ijkl}^{m}=0$ limit of (\ref{res-I4}) yields
\begin{equation}
\mathcal{I}_{4}=-\left( p^{0}\right) ^{2}q_{0}^{2}-2p^{0}q_{0}p^{i}q_{i}+%
\frac{1}{3}\left( 2q_{0}+q_{i}q_{j}\mathbf{d}^{ij}\right)
d_{klm}p^{k}p^{l}p^{m}-\frac{2}{3}p^{0}d^{ijk}q_{i}q_{j}q_{k},
\label{res-I4-symm}
\end{equation}
which is a manifestly $G_{5}$-invariant, alternative simple expression of $%
\mathcal{I}_{4}$, in $\mathcal{N}=2$ symmetric $d$-SK geometries, as well as
in all $d=4$ $\mathcal{N}>2$-extended supergravity theories whose scalar
manifold is characterised by a symmetric cubic geometry\footnote{%
With the exception of $\mathcal{N}=4$ \textit{``pure''} and of $\mathcal{N}%
=5 $ supergravities, these also are all $\mathcal{N}>2$-extended theories
which can be uplifted to $d=5$dimensions (see \textit{e.g.} \cite{LA08-Proc}
for quick reference Tables, and Refs. therein).}. In particular, for $%
G_{4}=E_{7\left( -25\right) }$ ($\mathcal{N}=2$, $d=4$ $J_{3}^{\mathbb{O}}$%
-based ``magic'' supergravity) and $G_{4}=E_{7\left( 7\right) }$ ($\mathcal{N%
}=8$, $d=4$ $J_{3}^{\mathbb{O}_{s}}$-based maximal supergravity), (\ref
{res-I4-symm}) provides an equivalent expression of the Cartan-Cremmer-Julia
\cite{Cartan,CJ} unique quartic invariant of the fundamental irrepr. $%
\mathbf{56}$ of the exceptional Lie group $E_{7}$. It is also worth
remarking that for symmetric $d$-SKG (and more in general in all cases in
which $E_{~ijkl}^{v}=0$ globally) the expressions (\ref{I4-bare-1})-(\ref
{defs-I4-bare}) and (\ref{res-I4-symm}) actually are scalar-independent and
thus purely charge-dependent, and therefore $\mathcal{I}_{4}$ actually is
the unique quartic invariant polynomial of the relevant symplectic (ir)repr.
$\mathbf{R}$ of the $d=4$ $U$-duality group $G_{4}$.

\item  The alternative expression (\ref{res-I4}) for $\mathcal{I}_{4}$ is
necessary to consistently match some known expressions of BH entropy with
the formalism of $d$-SK geometries. Concerning this, the $p^{0}=0$ limit of (%
\ref{res-I4-symm}) yields
\begin{equation}
\mathcal{I}_{4}=\frac{1}{3}\left( 2q_{0}+q_{i}q_{j}\mathbf{d}^{ij}\right)
d_{klm}p^{k}p^{l}p^{m},
\end{equation}
matching Eqs. (50)-(51) of \cite{TT1}. Actually, since the treatment of \cite
{TT1} deals with generic (not necessarily symmetric, nor homogeneous) $d$-SK
geometries, one should actually use the full formula (\ref{res-I4}).
Consequently, the consistence of the results (50)-(51) of \cite{TT1} with
the general formula (\ref{res-I4}) yields the following constraint on the
on-shell expression of the $E$-tensor (\textit{at least} for $p^{0}=0$):
\begin{equation}
\left. E_{~ijkl}^{m}\right| _{\partial V_{BH}=0}p^{j}p^{k}p^{l}q_{m}q_{n}%
\mathbf{d}^{in}=0.  \label{E-constr}
\end{equation}
It is known that the configuration $\left( p^{i},q_{0},q_{i}\right) $ does
not support axion-free attractor solutions \cite{CFM1}, thus (\ref{E-constr}%
) should be considered in an axionful background. However, the $E$-tensor is
insensitive to the presence of non-vanishing axions, because it only depends
on $\nu $ and $\widehat{\lambda }^{i}$'s, as given by Eq. (4.21) of \cite
{Raju-1}.

\item  The observations at point \textbf{1} are no more generally true in
\textit{non-symmetric} $d$-SK geometries, and in all cases in which the $E$%
-tensor does not vanish globally\footnote{%
For some elaborations on this issue, see \textit{e.g.} the recent treatment
given in \cite{Raju-1}.}. In this case, $\mathcal{I}_{4}$\textit{\ is no
more an invariant of the }$U$\textit{-duality group }$G_{4}$ (whose
transitive action on the scalar manifold is spoiled in the non-homogeneous
case; see \textit{e.g.} \cite{dWVVP}). Concerning this, it is worth
recalling that $G_{4}$ always contains (and for totally generic $d_{ijk}$'s,
coincides with) the semi-direct product of PQ axion-shifts (\ref{axion-shift}%
) $\mathbb{R}^{n_{V}}$ and an overall rescaling $SO\left( 1,1\right) $,
namely (see \textit{e.g.} \cite{dWVVP}):
\begin{equation}
SO\left( 1,1\right) \times _{s}\mathbb{R}^{n_{V}}\subset G_{4}.
\end{equation}
Within this framework, some analysis of the dependence on the scalar degrees
of freedom can be made. First of all, one can easily verify that in $d$-SK
geometries all relevant geometrical quantities considered above are
independent of the $d=4$ axions $x^{i}$, namely the real parts of the $d=4 $
complex scalars coordinatising the special K\"{a}hler vector multiplets'
scalar manifolds of $\mathcal{N}=2$, $d=4$ supergravity. This can ultimately
be traced back to the $d=5$ origin of all $d$-SK geometries, which are the
only SK geometries which can be uplifted to $5$ space-time dimensions. Then,
(\ref{a^ij-def})-(\ref{d^ijk-def}) and (\ref{a^ij-def}) yield that
\begin{eqnarray}
\frac{\partial \left( \nu ^{5/3}E_{~ijkl}^{v}\right) }{\partial \nu } &=&0;
\label{first-res} \\
&&  \notag \\
\nu ^{5/3}\frac{\partial E_{~ijkl}^{v}}{\partial \widehat{\lambda }^{v}} &=&%
\frac{3}{2^{5}}d_{p(ij}d_{kl)q}d_{rst}\left[ \frac{1}{2}\left( \delta
_{v}^{p}\widehat{\lambda }^{r}+\delta _{v}^{r}\widehat{\lambda }^{p}\right) +%
\widehat{d}^{pm}\widehat{d}^{rn}d_{vmn}\right] a^{qs}a^{vt}  \notag \\
&=&\frac{3}{2^{6}}\left[
\begin{array}{l}
d_{v(ij}d_{kl)q}d_{r}^{~vq}\widehat{\lambda }^{r}+d_{p(ij}d_{kl)q}d_{svt}%
\widehat{\lambda }^{p}a^{qs}a^{vt} \\
\\
+2d_{p(ij}d_{kl)q}d_{r}^{~vq}\widehat{d}^{pm}\widehat{d}^{rn}d_{vmn}
\end{array}
\right] .  \label{second-res}
\end{eqnarray}
The result (\ref{first-res}) was derived in \cite{Raju-1}. On the other
hand, (\ref{second-res}) expresses the way the $E$-tensor depends on $%
\widehat{\lambda }^{i}$'s, encoding the non-symmetric nature of the
corresponding $d$-SK geometry.
\end{enumerate}

\section*{Acknowledgments}

We would like to thank Sergio Ferrara for enlightening discussions.

The work of S. B. is supported by the ERC Advanced Grant no. 226455, \textit{%
``Supersymmetry, Quantum Gravity and Gauge Fields''} (\textit{SUPERFIELDS}).

The work of A. M. has been supported by an INFN visiting Theoretical
Fellowship at SITP, Stanford University, CA, USA. Furthermore, he would like
to thank Ms. Hanna Hacham for her nice and inspiring hospitality in Palo
Alto, CA, USA.

The work of R. R. has been supported in part by Dipartimento di Scienza
Fisiche of ``Federico II'' University and INFN-Sez. di Napoli. He would also
like to thank INFN - LNF for kind hospitality and support.

\end{document}